\documentclass{aa}
\usepackage[]{natbib}
\usepackage{psfig,graphicx,ulem,epsfig}
\usepackage{txfonts}
\usepackage{soul,color}
\usepackage{longtable}

\def\mathnew{\mathsurround=0pt}
\def\simov#1#2{\lower .5pt\vbox{\baselineskip0pt \lineskip-.5pt
\ialign{$\mathnew#1\hfil##\hfil$\crcr#2\crcr\sim\crcr}}}

\begin{document}

\title{The merging cluster Abell~1758 revisited: multi-wavelength
  observations and numerical simulations\thanks{Based on archive data
    retrieved from the Canadian Astronomy Data Centre Megapipe archive
    and obtained with MegaPrime/MegaCam, a joint project of CFHT and
    CEA/DAPNIA, at the Canada-France-Hawaii Telescope (CFHT) which is
    operated by the National Research Council (NRC) of Canada, the
    Institut National des Sciences de l'Univers of the Centre National
    de la Recherche Scientifique (CNRS) of France, and the University
    of Hawaii. The X-ray analysis is based on XMM-\textit{Newton}
    archive data.  This research has made use of the NASA/IPAC
    Extragalactic Database (NED) which is operated by the Jet
    Propulsion Laboratory, California Institute of Technology, under
    contract with the National Aeronautics and Space Administration,
    and of the SIMBAD database, operated at CDS, Strasbourg, France.}}

\author{
F.~Durret \inst{1,2} \and
T.~F.~Lagan\'a \inst{3} \and
M.~Haider \inst{4} 
}

\institute{
UPMC Universit\'e Paris 06, UMR~7095, Institut d'Astrophysique de Paris, 
F-75014, Paris, France
\and
CNRS, UMR~7095, Institut d'Astrophysique de Paris, F-75014, Paris, France
\and
IAG, USP, R. do Mat\~ao 1226, 05508-090, S\~ao Paulo/SP, Brazil
\and
Inst. of Astro and Particle Physics, University of Innsbruck, A-6020 Innsbruck, Austria
}

\date{Accepted . Received ; Draft printed: \today}

\authorrunning{Durret, Lagan\'a \& Haider}

\titlerunning{Abell 1758}

\abstract
{We discuss here the properties of the double cluster Abell~1758, at
  a redshift z$\sim $0.279, which shows strong evidence for merging.}
{We analyse the optical properties of the North and South clusters
 of Abell~1758 based on deep imaging obtained with the CFHT archive
  Megaprime/Megacam camera in the $g^\prime$ and $r^\prime$ bands, covering a
  total region of about 1.05$\times$1.16~deg$^2$, or $16.1\times
  17.6$~Mpc$^2$. Our X-ray analysis is based on archive
  XMM-\textit{Newton} images. Numerical simulations were performed
  using an N-body algorithm to treat the dark matter component, a
  semi-analytical galaxy formation model for the evolution of the
  galaxies and a grid-based hydrodynamic code with a PPM scheme for
  the dynamics of the intra-cluster medium. We have computed galaxy
  luminosity functions (GLFs) and 2D temperature and metallicity maps
  of the X-ray gas, which we then compared to the results of our
  numerical simulations.}
{The GLFs of Abell~1758 North are well fit by Schechter functions in
  the $g'$ and $r'$ bands, but with a small excess of bright galaxies,
  particularly in the $r'$ band; their faint end slopes are similar in
  both bands. On the contrary, the GLFs of Abell~1758 South are not
  well fit by Schechter functions: excesses of bright galaxies are
  seen in both bands; the faint end of the GLF is not very well
  defined in $g'$.  The GLF computed from our numerical simulations
  assuming a halo mass--luminosity relation agrees with those derived
  from the observations.  From the X-ray analysis, the most striking
  features are structures in the metal distribution.  We found two
  elongated regions of high metallicity in Abell~1758 North with two
  peaks towards the center. On the other hand, Abell~1758 South shows
  a deficit of metals in its central regions. Comparing observational
  results to those derived from numerical simulations, we could mimic
  the most prominent features present in the metallicity map and
  propose an explanation for the dynamical history of the cluster. We
  found in particular that in the metal rich elongated regions of the
  North cluster, winds had been more efficient in transporting metal
  enriched gas to the outskirts than ram pressure stripping.}
 {We confirm the merging structure of each of the North and South
  clusters, both at optical and X-ray wavelengths.
}


\maketitle

\section{Introduction}

Environmental effects are known to have an influence on galaxy
evolution, and can therefore modify Galaxy Luminosity Functions
(hereafter GLFs). This is particularly obvious in merging clusters,
where GLFs may differ from those in non-merging (relaxed) clusters,
and where GLFs may also be observed to differ between one photometric
band and another \citep[e.g.][]{boue08} and references
therein). GLFs also allow to trace the cluster formation history, as
shown for example in the case of Coma \citep{adami07}.  

This dynamical history can also be derived by analyzing the
temperature and metallicity distributions of the X-ray gas in
clusters.  Such maps have revealed that in many cases, clusters with
emissivity maps showing a rather relaxed appearance could have very
disturbed temperature and metallicity distributions \citep[see
e.g.][~and references therein]{durret08}, meaning that they have
undergone one or several mergers in the last few Gyrs. The study of
the thermal structure of the intra-cluster medium (ICM) indeed
provides a very interesting record of the dynamical processes that
clusters of galaxies have experienced during their formation and
evolution.  The temperature distribution of the ICM gives us insight
into the process of galaxy cluster merging and on the dissipation of
the merger energy in form of turbulent motion.  Metallicity maps can
indeed be regarded as a record of the integral yield of all the
different stars that have released their metals through supernova
explosions or winds during cluster evolution.

The comparison of temperature and metallicity maps to the results of
hydrodynamical numerical simulations allow to characterize the last
merging events which have taken or are taking place. For example, the
comparison of the complex temperature and metallicity maps of Abell~85
\cite{durret05} with the numerical simulations by \citet{bourdin04} 
show that two or three mergers have taken place at various
epochs in this cluster in the last few Gyrs, besides the ongoing
merger seen as a filament made of groups falling on to the cluster
\citep{durret03}.

We have become interested in pairs of clusters, where the effects of
merging are expected to be even stronger. In some cases, one of the
clusters shows itself a double structure (Abell~223, Abell~1758
North). By coupling deep optical multi-band imaging with X-ray maps,
we have recently analyzed the Abell~222/223 cluster pair \citep{durret10}. 
We found that Abell~222 (the less perturbed and less
massive cluster) had GLFs well fit by a Schechter function, with a
steeper faint end in the $r'$ band than in the $g'$ band, implying
little star formation; its X-ray gas showed quite homogeneous
temperature and metallicity maps, but with no cool core, suggesting
that some kind of merger must have taken place to suppress the cool
core. This was confirmed by the distribution of bright galaxies in this
cluster, which also suggests that this cluster is not fully
relaxed. Abell~223 (the most perturbed and massive cluster), was found
to have comparable GLFs in both bands, with an excess of galaxies over
a Schechter function at bright magnitudes.  Its temperature and
metallicity distributions were found to be very inhomogeneous,
implying that it has most probably just been crossed by a smaller
cluster, which now appears at the north east tip of the maps. Note
that a bridge of galaxies seems to exist between the two clusters
\citep{dietrich02}, as well as a possible dark matter filament
joining the two clusters \citet{dietrich05}.

The Abell~1758 cluster, at a redshift of 0.279, was analysed in X-rays
by \citet{DK04} based on Chandra and XMM-\textit{Newton}
data. These authors showed that this cluster is in fact double, with a
North and a South component separated by approximately 8~arcmin (2~Mpc
in projection), both undergoing major mergers, with evidence for X-ray
emission between the two clusters. However, very little has been
published on this system at optical wavelengths, and few galaxy
redshifts are available. As a second study of cluster pairs, we chose
to analyse this system by coupling archive optical CFHT Megacam data
with archive XMM-\textit{Newton} data. This allowed us to compute GLFs
in two bands as well as temperature and metallicy maps for the ICM,
which were compared to the results of numerical simulations.

For a redshift of 0.279, Ned Wright's cosmology calculator\footnote{http://nedwww.ipac.caltech.edu/} \citep{wright06}
gives a luminosity distance of
1428~Mpc and a spatial scale of 4.233~kpc/arcsec, giving a distance
modulus of 40.77 (assuming a flat $\Lambda$CDM cosmology with
H$_0=70$~km~s$^{-1}$~Mpc$^{-1}$, $\Omega _M=0.3$ and $\Omega _\Lambda
=0.7$).

The paper is organised as follows. We describe our optical analysis in
Section 2, and results concerning the observed and simulated galaxy
luminosity functions in Section 3.  The X-ray data analysis and
results, including temperature and metallicity maps, are presented in
Section~4.  The results of numerical simulations that were run to help
us to account for the X-ray temperature and metallicity maps are
described in Section~5. An overall picture of this cluster pair is
drawn in Section~6.

\section{Optical data and analysis}

\subsection{The optical data}

We have retrieved from the CADC Megapipe archive \citep{gwyn09} the
reduced and stacked images in the $g'$ and $r'$ bands (namely
G008.203.140+50.518.G.fits and G008.203.140+50.518.R.fits) and give a few
details on the observations in Table~\ref{tab:obs}.  Observations were
made at the CFHT with the Megaprime/Megacam camera, which has a pixel
size of $0.186\times 0.186$~arcsec$^2$.

\begin{table}[h!] 
\caption{Summary of the observations.}
\begin{center}
\begin{tabular}{lrr}
\hline
\hline
Filter          & $g'$    & $r'$ \\     
\hline
Number of coadded images & 4   & 9 \\
Total exposure time (s)   & 1800  & 4860  \\
Seeing (arcsec)     & 0.75  & 0.65 \\
Limiting magnitude (5$\sigma$) & 27.1 & 26.8 \\
\hline
\end{tabular}
\end{center}
\label{tab:obs}
\end{table}

We did not use the catalogues available for these images, because they
were made without masking the surroundings of bright stars, so we
preferred to build masks first, then to extract sources with
SExtractor \citep{bertin96}.  The total area covered by the
images was 20403$\times$22406~pixels$^2$, or 1.05$\times$1.16~deg$^2$
($16.1\times 17.6$~Mpc$^2$ at the cluster redshift).

Objects were detected and measured in the full $r'$ image, then
measured in the $g'$ image in double image mode (i.e. the objects
detected in $r'$ were then measured in $g'$ exactly in the same way as
in $r'$). Magnitudes are in the AB system. The objects located in the
masked regions were then taken out of the catalogue, leading to a
final catalogue of 298,170 objects.  We created masks around bright
stars and image defects in the portion of the image covered by
Abell~1758 and by the ring around the cluster that was used to
estimate the background galaxy contamination to the GLF (due to its
relatively high redshift the cluster does not cover the entire image).
After taking out masked objects we were left with a catalogue of
286,505 objects with measured $r'$ magnitudes, out of which 277,646
also have measured $g'$ magnitudes.

Since the seeing was better in the $r'$ band (see
Table~\ref{tab:obs}), we performed our star-galaxy separation in this
band.

\subsection{Star-galaxy separation}
\label{stargalsep}
\begin{figure} 
\centering \mbox{\psfig{figure=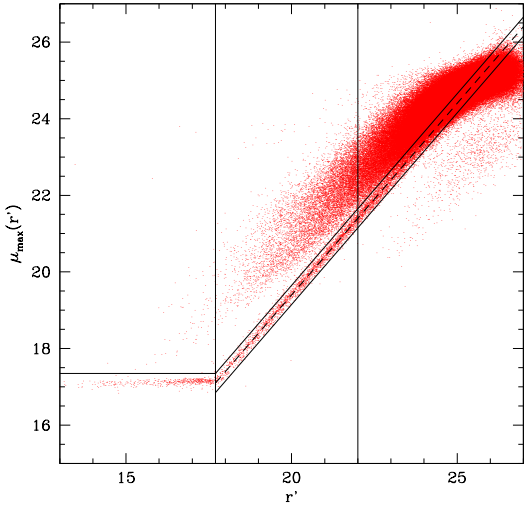,width=8cm}}
\caption   {Central surface brightness in the $r'$ band as a function
  of $r'$ magnitude. The horizontal and oblique full lines isolate the
  star sequence (below the horizontal line and between the oblique
  lines).  The two vertical lines correspond to $r'=17.75$ where the
  stars stop being saturated, and $r'=22$, where the fit to calculate
  the star sequence was limited (see text).}
\label{fig:r_mumax}
\end{figure}

In order to separate stars from galaxies, we plotted the maximum
surface brightness $\mu_{max}$ in the $r'$ band as a function of
$r'$. The result is shown in Fig.~\ref{fig:r_mumax}.

The best fit to the star sequence visible on Fig.~\ref{fig:r_mumax}
calculated for $17.75<r'<22$ is $\mu_{max}=0.996r'-0.524$, with
standard deviations on the slope and constant of 0.002 and 0.035
respectively. The point-source (hereafter called ``star'') sequence is
clearly visible for $r'<22$, with the star saturation showing well for
$r'<17.75$.  We will define galaxies as the objects with $\mu
_{max}(r')>17.4$ for $r'<17.75$, and as the objects above the line of
equation $\mu_{max}=0.996r'-0.274$ for $r'\geq 17.75$.  Stars will be
defined as all the other objects (see Fig.~\ref{fig:r_mumax}).  The
small cloud of points observed in Fig.~\ref{fig:r_mumax} under the
star sequence is in fact defects, but represents less than 2\% of the
number of stars.  We thus obtained a star and a galaxy catalogue.

As a check to see up to what magnitude we could trust our star-galaxy
separation, we retrieved the star catalogue from the Besan\c{c}on
model for our Galaxy (Robin et al. 2003) in a 1~deg$^2$ region
centered on the position of the image analysed here. Such a catalogue
is in AB magnitudes (as ours) and is corrected for extinction. In
order for it to be directly comparable to our star catalogue, we
corrected our star catalogue (and our galaxy catalogue as well, for
later purposes) for extinction: 0.0531~mag in $g'$ and 0.0385~mag in
$r'$ (as derived from the Schlegel et al. 1998 maps).

The $r'$ magnitude histogram of the objects classified as stars in our
$r'$ image roughly agrees with the Besan\c con star catalogue for
$r'\leq$22. However, for $r'>$21, we start to detect more stars than
predicted by the Besan\c con model (the difference is only about 15\%
at $r'=21.5$ but becomes 25\% in the $r'=22$ bin, and the
difference continues to increase at fainter magnitudes). 

We will therefore consider that our star-galaxy separation is correct
for $r'\leq 22$. For fainter magnitudes, we will compute galaxy counts by
counting the total number of objects (galaxies plus stars) per bin of
0.5~mag, and considering that the number of galaxies is equal to the
total number of objects minus the number of stars predicted in each
bin by the Besan\c con model.

\begin{figure} 
\centering \mbox{\psfig{figure=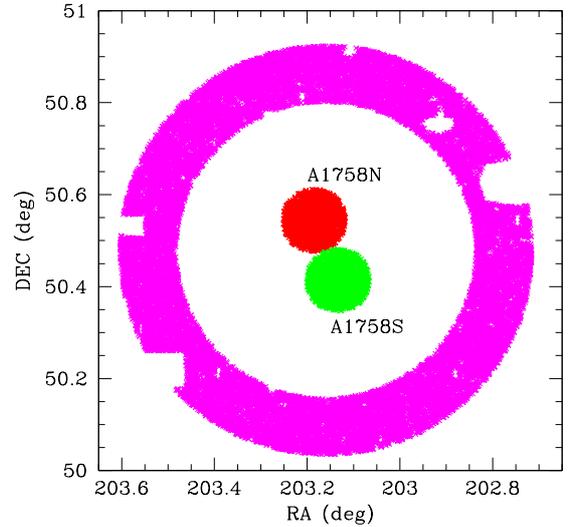,width=8cm}}
\caption   {Positions of the objects (stars and galaxies) in the
  Abell~1758 North (red) and South (green) catalogues. The objects
  used to estimate background counts (see text) are shown in magenta.
  Note that the figure covers $1\times 1$~deg$^2$, and is slightly
  smaller than the images.}
\label{fig:xy}
\end{figure}

We extracted from the star and galaxy catalogues two catalogues as
large as possible corresponding to the North and South clusters.
Their respective positions were taken to be the X-ray centers of the
two clusters, as derived from XMM-\textit{Newton} data:
203.1851,+50.5445 (J2000.0, in degrees) for the North cluster, and
203.1335,+50.4138 for the South one.  The maximum possible radius to
obtain independent catalogues for each of the two clusters was
0.0675~deg, or 1.03~Mpc at a redshift of 0.279.

We obtained for each of the two clusters three complementary
catalogues (with $g'$ and $r'$ magnitudes): objects classified as
galaxies (classification valid at least for $r'\leq 22$), objects
classified as stars, and a complete catalogue of galaxies+stars which
will be used for $r'>22$. The positions of the galaxies in the regions
of the two clusters are shown in Fig.~\ref{fig:xy}.

\subsection{Catalogue completeness}

\begin{figure} \centering
\mbox{\psfig{figure=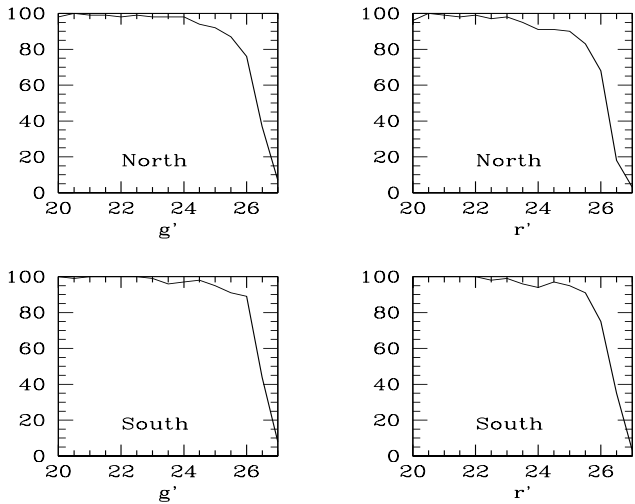,width=9cm,height=7cm,angle=0}}
\caption   {Point source completeness as a function of magnitude in
  percentages for Abell~1758 North (top) and South (bottom) in $g'$
  (left) and $r'$ (right) for point--like objects (see
  text). }  \label{fig:compl}
\end{figure}

The completeness of the catalogue is estimated by simulations.  For
this, we add ``artificial stars'' (i.e. 2-dimensional Gaussian
profiles with the same Full-Width at Half Maximum as the average image
Point Spread Function) of different magnitudes to the CCD images and
then attempt to recover them by running SExtractor again with the same
parameters used for object detection and classification on the
original images. In this way, the completeness is measured on the
original images.

In practice, we extract from the full field of view two subimages,
each $1300\times 1300$~pixels$^2$, corresponding to the positions of
the two clusters on the image.

In each subfield, and for each 0.5 magnitude bin between $r'=20$ and 27,
we generate and add to the image one star that we then try to detect
with SExtractor, assuming the same parameters as previously. This
process is repeated 100 times for each of the two fields and bands.

Such simulations give a completeness percentage for stars. This is
obviously an upper limit for the completeness level for galaxies,
since stars are easier to detect than galaxies. However, we have shown
in a previous paper that this method gives a good estimate of the
completeness for normal galaxies if we apply a shift of $\sim 0.5$~mag
\citep[see][]{adami06}. Results are shown in Fig.~\ref{fig:compl}.

From these simulations, and taking into account the fact that results
are worse by $\sim 0.5$~mag for mean galaxy populations than for
stars, we can consider that our galaxy catalogues are complete to
better than 80\% for $g'\leq 25.8$ and $r'\leq 25.6$ in both clusters.

\subsection{Galaxy counts}

The surfaces covered by the Abell~1758 North and South catalogues (after
excluding masked regions) are 0.01412~deg$^2$ and 0.01418~deg$^2$
respectively. Galaxy counts were computed in bins of 0.5~mag
normalized to a surface of 1~deg$^2$.

For $r' \leq22$, galaxy counts were derived directly by computing
histograms of the numbers of galaxies in the Abell~1758 North and South
catalogues.  For $r'>22$, we built for each cluster histograms of the
total numbers of objects (galaxies+stars) and obtained galaxy counts
by subtracting the numbers of stars predicted by the Besan\c con
model.  The resulting galaxy counts will be used in the next section
to derive the GLFs for both clusters in both bands.

As a test, we considered the galaxy counts in the 21.5--22.0 magnitude
bin computed for both clusters with the two methods (i.e.  first
method: considering that the star-galaxy separation is valid, and
second method: considering the total number of objects
(galaxies+stars) and subtracting the number of stars predicted by the
Besan\c con model to obtain the number of galaxies).  In all cases,
the differences are smaller than 4\%. 
 
Note that no k-correction was applied to the galaxy magnitudes.

\section{Results: colour-magnitude diagrams and galaxy luminosity
  functions}
\label{sec:GLF}

In order to compute the galaxy luminosity functions of the two
clusters, we need to subtract to the total galaxy counts the number
counts corresponding to the contamination by the foreground and
background galaxies.

For galaxies brighter than $r'=22$ we will select galaxies with a high
probability to belong to the clusters by drawing colour-magnitude
diagrams and selecting galaxies located close to this relation.  A few
spirals may be missed in this way, but their number in any case is
expected to be small, as explained in Sect.~3.1 (also see
e.g. Adami et al. 1998).  For galaxies fainter than $r\prime=22$ we will
subtract galaxy counts statistically.

\subsection{Colour--magnitude diagram}

\begin{figure} 
\centering \mbox{\psfig{figure=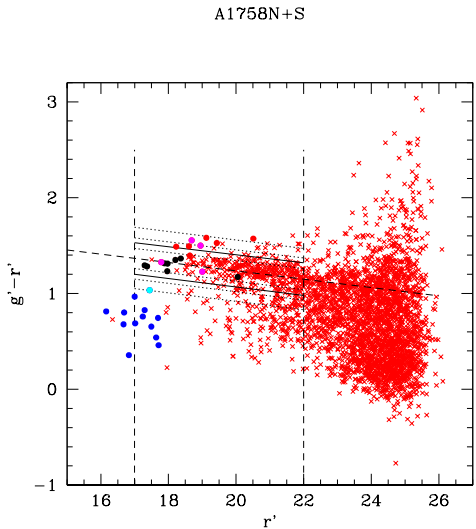,width=7cm}}
\caption   {$(g\prime-r\prime)$ vs. $r'$ colour--magnitude diagram for Abell~1758
  (North and South) for objects classified as galaxies from the
  $\mu_{max} - r\prime$ magnitude relation. The vertical dashed lines indicate
  the magnitude interval where the colour-magnitude relation was
  computed. The long oblique dashed line shows the mean
  colour-magnitude relation; the short oblique dotted lines indicate
  intervals of $\pm 1\sigma , 2\sigma$ and $3\sigma$ around the
  colour-magnitude relation.  The interval finally adopted of
  $1.5\sigma$ is shown with full lines.  The filled circles show
  the galaxies with measured redshifts, colour--coded as follows:
  black: galaxies belonging to the cluster according to their
  spectroscopic redshifts, cyan and magenta for galaxies with smaller
  and larger redshifts than the cluster, but inside the two circles
  where the two clusters were extracted, blue and red: galaxies
  with smaller and larger redshifts than the cluster, but outside
  these two circles. }
\label{fig:coulmag}
\end{figure}

The $g'-r'$ vs. $r'$ colour--magnitude diagram is shown in
Fig.~\ref{fig:coulmag} for the two clusters together (the diagrams
are the same for both clusters). A sequence is well defined for
galaxies in the magnitude range $17<r'<22$ in both clusters. We
computed the best fit to the $g'-r'$ vs. $r'$ relations in this
magnitude range by applying a linear regression.  We then eliminated
the galaxies located more than 3$\sigma$ away from this relation and
computed the $g'-r'$ vs. $r'$ relation again.

The equation of the colour--magnitude relation is found to be:
$g'-r'=-0.0436r'+2.108$ with an r.m.s. on the constant $\sigma=0.11$.

We also plot in Fig.~\ref{fig:coulmag} galaxies with measured
spectroscopic redshifts. We can see that the positions of the galaxies
belonging to the cluster according to their spectroscopic redshifts,
i.e.  with redshifts in the  [0.264, 0.294]   interval fall very close to
the best fit to the colour--magnitude relation.

In view of this, for $r'\leq 22$, we will consider hereafter that all the
galaxies located within $\pm 1.5\sigma$ of the colour--magnitude relation
(i.e. between the two black lines of Fig.~\ref{fig:coulmag}) belong to
the cluster. 

We can note that the scatter $\sigma=0.11$ is somewhat larger than
found in the Abell~222/223 clusters at z=0.21 (Durret et al. 2010),
but the interval chosen for cluster membership ($\pm 1.5\sigma$)
remains smaller than that used for example by Lagan\'a et al. (2010)
in the redshift interval [0.11,0.23].

The initial sample of galaxies comprises 1599 and 1438 galaxies in the
North and South clusters respectively (with no magnitude
limit). Within the $\pm 1.5\sigma$ interval along the red sequence, we
are left with respective numbers of galaxies of 477 and 340. These
numbers become 192 and 116 galaxies for $r'\leq 22$. The North cluster
is obviously richer than the South one.

With this rather strict criterium, we obviously select galaxies with a
high probability to belong to the cluster, but we may lose some
galaxies, in particular blue cluster galaxies falling under the
sequence.

We have therefore estimated the number of blue cluster galaxies lost
by selecting galaxies within $\pm 1.5 \sigma$ of the red sequence in the
following way. First, we computed histograms of numbers of galaxies
within $\pm 1.5\sigma$ of the red sequence and below this sequence in
bins of 1 absolute magnitude in the $r'$ band. These counts were made
in absolute magnitude bins to be comparable to the counts estimated
from luminosity functions of field galaxies.  The bins of interest
here are between ${\rm M_{r'}=-22}$ and $-19$, roughly corresponding
to $r'=18.5$ and 21.5.  We then computed the number of foreground
galaxies expected.  Since the comoving volume at z=0.279 is
5.834~Gpc$^3$ (Wright 2006), and each of our clusters covers an
area of 0.01431~deg$^2$ on the sky, the volume in the direction of
each cluster is 2024~Mpc$^3$. By using the R band luminosity
function by Ilbert et al. (2005) in the  0.05-0.20   redshift bin (see
their Fig. 6 and Table 1), we find that the percentages of ``lost''
galaxies are of the order of 30\% for ${\rm M}_{r'}=-19$, of
10\%--15\% for ${\rm M}_{r'}=-20$ and $-21$, and less than 10\% for
brighter galaxies.

\subsection{Comparison field}
\label{sec:compfield}

\begin{figure} 
\centering \mbox{\psfig{figure=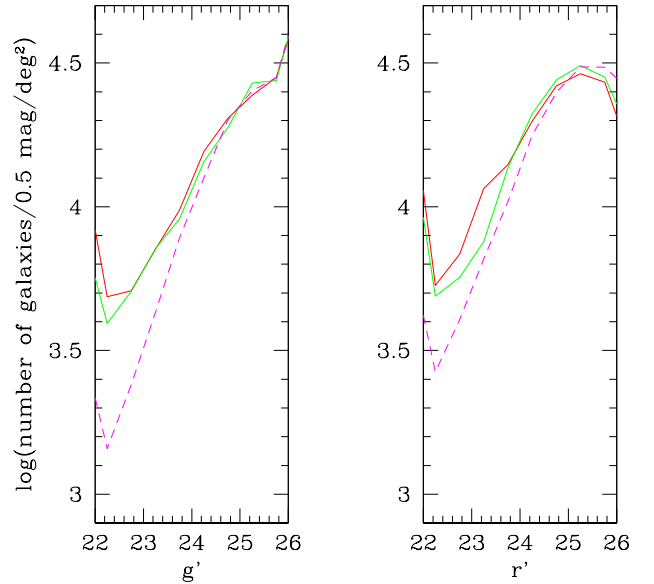,width=9cm}}
\caption   {Galaxy counts in the $g'$ (left) and $r'$ (right) bands for
  magnitudes $r'>22$ where the background must be subtracted
  statistically, in logarithmic scale.  The counts in Abell~1758 North
  and South are drawn in red and green respectively. The magenta
  dashed lines show the galaxy counts from the ``local'' background
  extracted in the annulus shown in Fig.~\ref{fig:xy}. Error bars are
  Poissonian and are not plotted for clarity.}
\label{fig:cts_bkgd}
\end{figure}

In order to perform a statistical subtraction of the background
contribution for $r'>22$, and since the cluster is quite distant and
does not cover the whole field, we extracted background counts in an
annulus surrounding the clusters (see Fig.~\ref{fig:xy}). The annulus
was centered on the middle position between the two clusters
(203.1593,+50.4792 J2000.0 in degrees), with an inner radius of
0.3232~deg (4.9~Mpc), and an outer radius of 0.4444~deg (6.8~Mpc).
The unmasked surface of the annulus is 0.278 deg$^2$. 

We checked if there could be any contamination of background counts in
this annulus by a group or cluster, and found a structure
west-southwest of Abell~1758. This appears as a cluster in 
Simbad\footnote{http://simbad.u-strasbg.fr/simbad/} 
with coordinates 13$^h$34$^{mn}$45.41$^s$, +50$^\circ$26'01.4'' (J2000.0) and redshift 0.085. The mean redshift for the 20
galaxies extracted from NED (http://nedwww.ipac.caltech.edu/) in this
region is 0.0869, with a dispersion in redshift $\sigma _z=0.0015$. This
cluster falls just outside the annulus which was used to estimate the
background counts subtracted to the galaxy counts to compute the GLF,
so its presence should not modify galaxy counts inside the annulus.

We can note from Fig.~\ref{fig:cts_bkgd} that galaxy counts are
comparable in Abell~1758 North and South in the $g'$ band, but differ
in the $r'$ band, where Abell~1758 North has more galaxies in the
$22\leq r'\leq 24$ magnitude range.

\subsection{Galaxy luminosity functions}
\label{sec:glf}

The Galaxy Luminosity Functions (GLFs) of Abell~1758 North and South
were calculated in bins of 0.5~mag and normalized to 1~deg$^2$. We
subtracted the background contribution using as background galaxy
counts the ``local'' counts in the annulus shown in magenta in
Fig.~\ref{fig:xy}.

\begin{figure} 
\centering \mbox{\psfig{figure=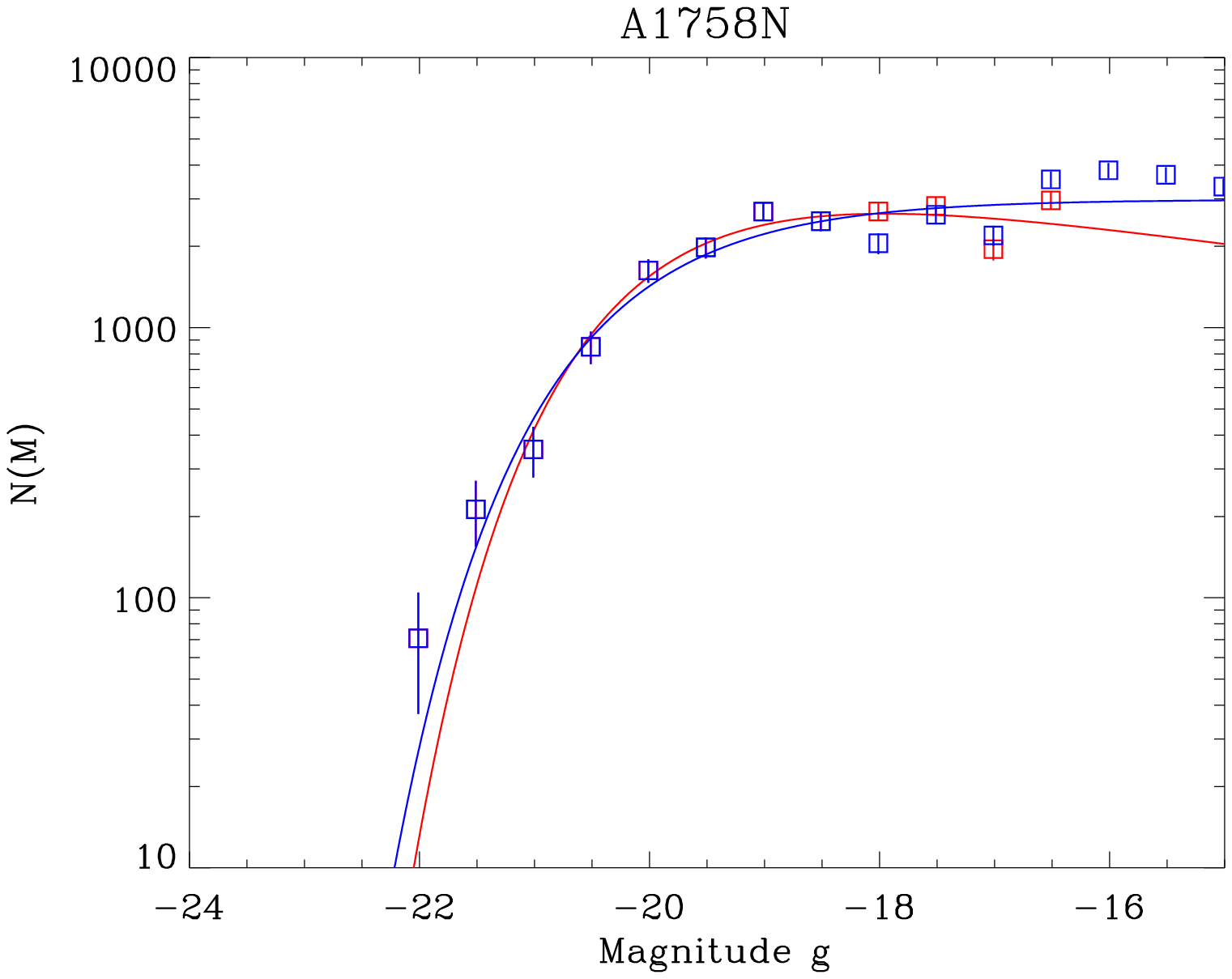,width=8cm}}
\centering \mbox{\psfig{figure=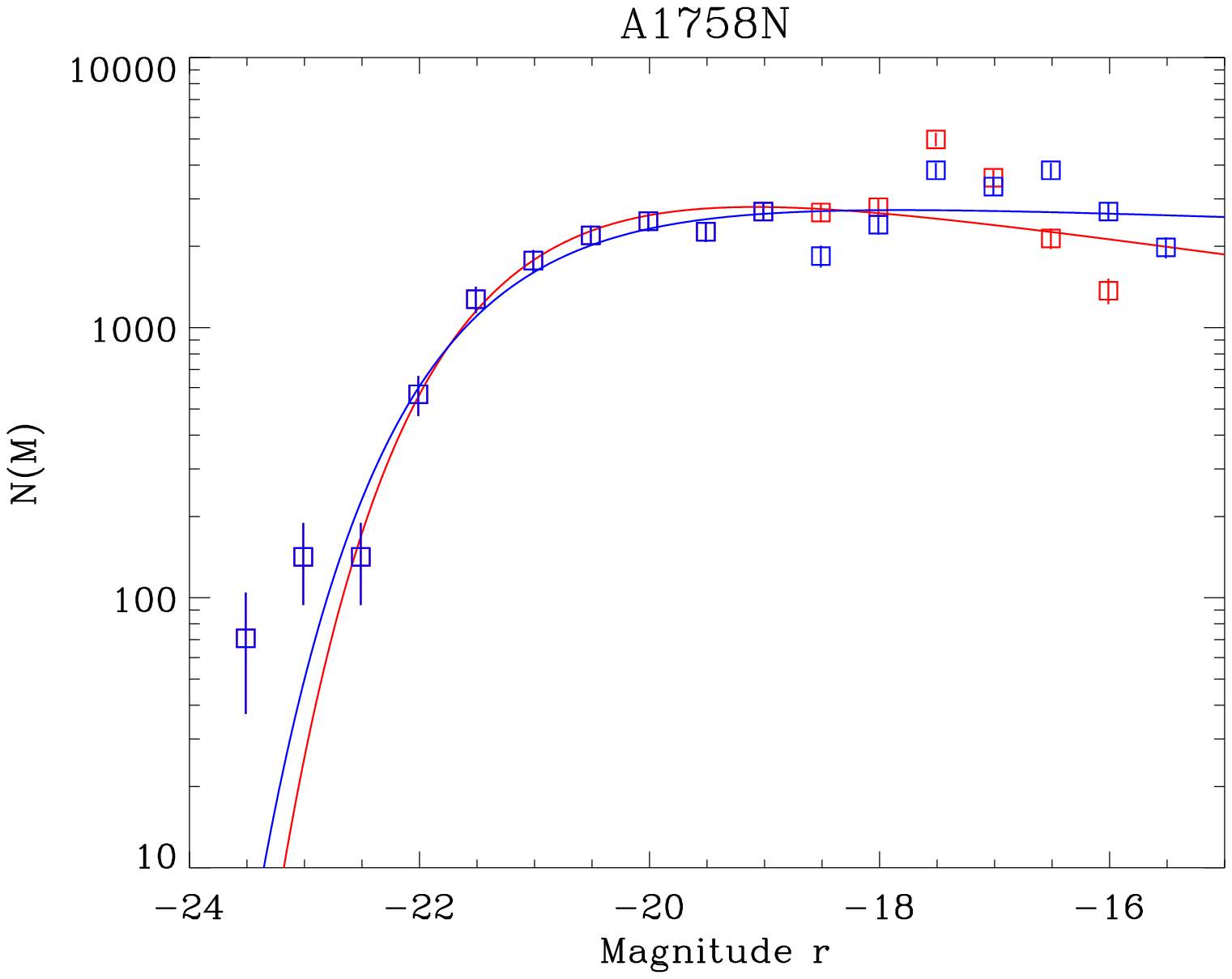,width=8cm}}
\caption {Galaxy luminosity functions for Abell~1758 North in the $g'$
  (top) and $r'$ (bottom) bands, in logarithmic scale. The blue and
  red points correspond to the two galaxy selections (see text), and
  the best Schechter function fits are drawn with the same colours as
  the corresponding points. Note that at bright magnitudes, the points
  exactly coincide, and since the blue points were plotted after the
  red ones, they appear blue.  Error bars are 4 times the Poissonian
  errors on galaxy counts (see text). }
\label{fig:GLF_A1758N}
\end{figure}

\begin{figure} 
\centering \mbox{\psfig{figure=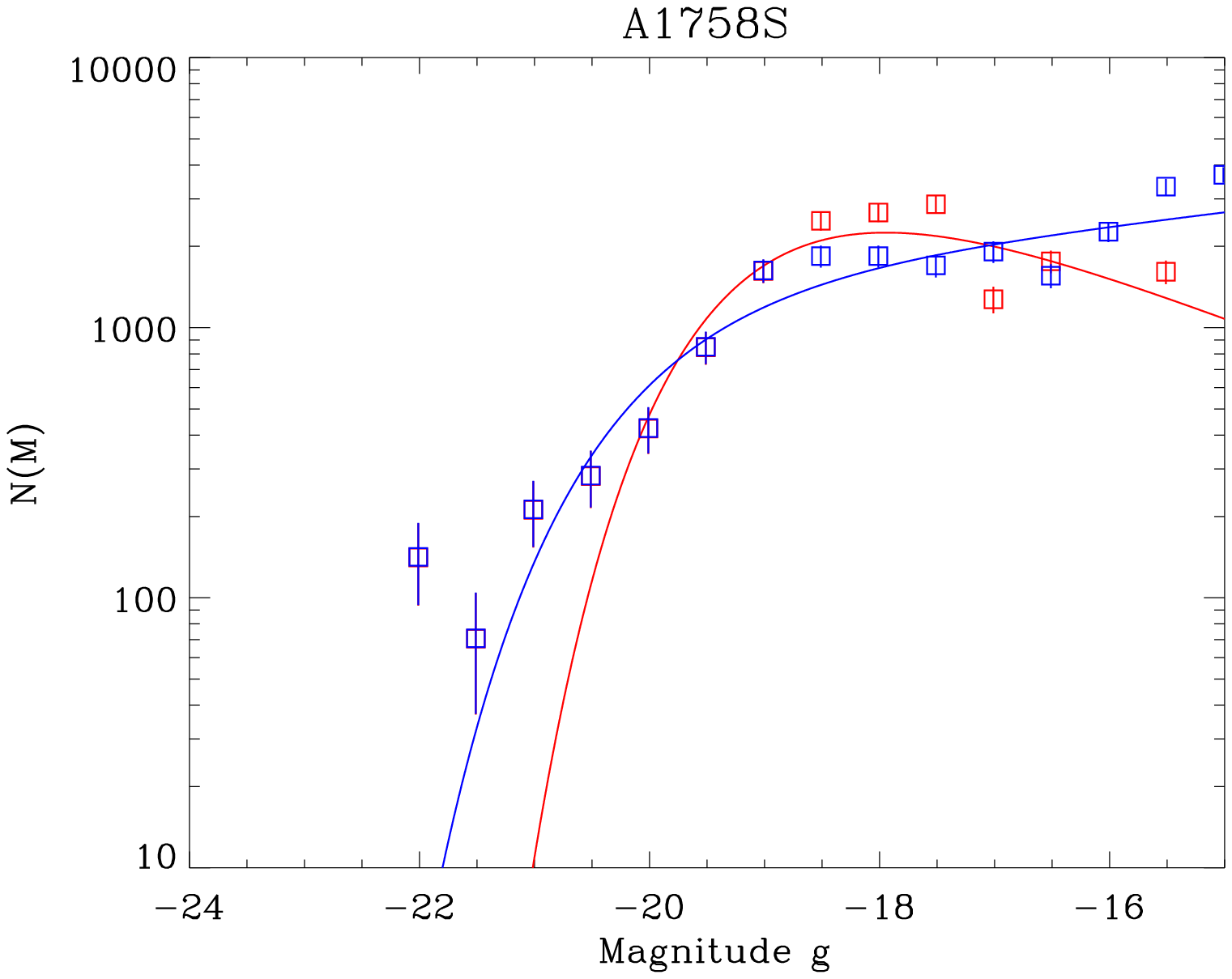,width=8cm}}
\centering \mbox{\psfig{figure=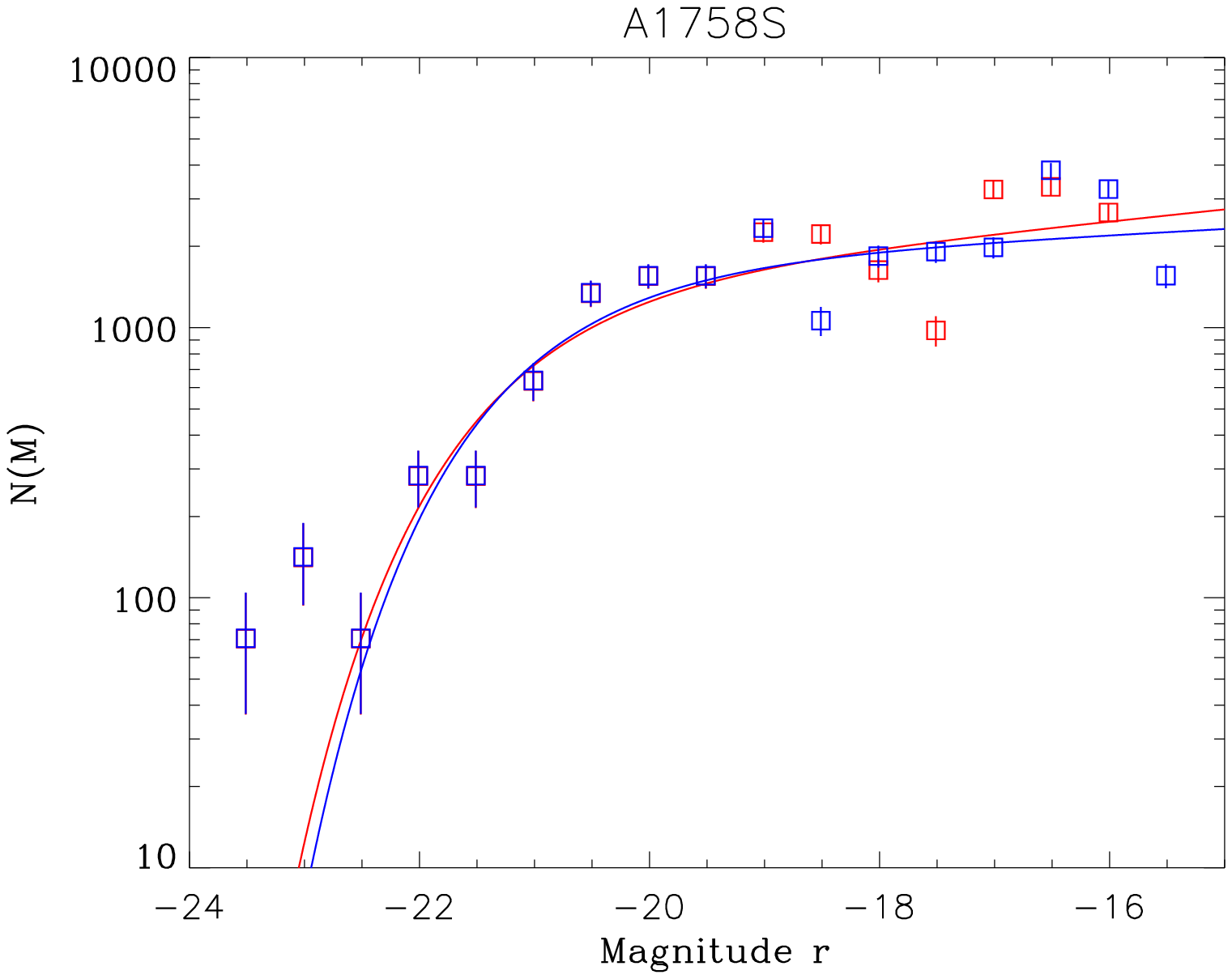,width=8cm}}
\caption   {Same as Fig.~\ref{fig:GLF_A1758N} for the South cluster.}
\label{fig:GLF_A1758S}
\end{figure}

The GLFs are displayed for Abell~1758 North and South in
Figs.~\ref{fig:GLF_A1758N} and \ref{fig:GLF_A1758S} respectively (red
points).  The error bars drawn in these figures were taken to be 4
times the Poissonian errors on galaxy counts, as derived from detailed
simulations previously performed by our team for similar data
\citep[see][~Fig.~5]{boue08}.

The GLFs (as a function of absolute magnitude) were fit by a Schechter function:

$$ S(M) = 0.4 \, \ln 10 \, \phi^{\ast} \, y^{\alpha+1} \, e^{-y} $$
with 
$y=10^{0.4 \, (M^{\ast}-M)} $.

The parameters of the Schechter function fits of the GLFs are given in
Table~\ref{tab:schechter}. The absolute magnitude ranges considered
are indicated for each fit, and GLFs and their fits are drawn in
Figs.~\ref{fig:GLF_A1758N} and \ref{fig:GLF_A1758S}.

\begin{table*}
  \caption{Schechter parameters for galaxy luminosity functions. The first 
    set of fits corresponds to galaxies selected from the colour-magnitude 
    relation for $r'<22$ and to a statistical background subtraction with 
    galaxy counts taken from the CFHTLS Deep field counts for $r'>22$. The 
    second set of fits corresponds to galaxies selected from the 
    colour-magnitude relation at all magnitudes. The last line corresponds 
    to the fit to the simulated GLF (see text).}
\begin{center}
\begin{tabular}{lccccc}
\hline
\hline
Cluster& Filter & Range & $\Phi ^*$ & M$^*$   & $\alpha$   \\     
\hline
North & $g'$ & $ -22.0,-16.5  $ & $4384 \pm 462 $ & $-20.06 \pm 0.12$ & $-0.86 \pm 0.05$ \\
      & $r'$ & $ -23.5,-16.0  $ & $4650 \pm 286 $ & $-21.18 \pm 0.08$ & $-0.85 \pm 0.02$ \\
South & $g'$ & $ -22.0,-15.5  $ & $5319 \pm 574 $ & $-18.90 \pm 0.14$ & $-0.59 \pm 0.08$ \\
      & $r'$ & $ -23.5,-16.0  $ & $1582 \pm 229 $ & $-21.34 \pm 0.16$ & $-1.11 \pm 0.04$ \\
\hline
North & $g'$ & $[-22.0,-16.5]$ & $3244 \pm 236 $ & $-20.33 \pm 0.09$ & $-1.00 \pm 0.02$ \\
      & $r'$ & $[-23.5,-16.0]$ & $3451 \pm 229 $ & $-21.44 \pm 0.08$ & $-0.96 \pm 0.02$ \\
South & $g'$ & $[-22.0,-15.5]$ & $1652 \pm 205 $ & $-20.09 \pm 0.16$ & $-1.12 \pm 0.03$ \\
      & $r'$ & $[-23.5,-16.0]$ & $1849 \pm 195 $ & $-21.18 \pm 0.14$ & $-1.06 \pm 0.03$ \\
\hline
Simulated &  & $[-25.0,-15.0]$ & $755 \pm 21$ & $-22.09 \pm 0.04$ & $-1.01 \pm 0.01$ \\
\hline
\end{tabular}
\end{center}
\label{tab:schechter}
\end{table*}

If we look at Abell~1758 North (Fig.~\ref{fig:GLF_A1758N}) we see that
a Schechter function fits rather well most of the GLF points for both
bands. However, there is an excess of galaxies over a Schechter
function in the very brightest magnitude bins, specially in the $r'$
band. There is also a ``bump'' around $r'$ absolute magnitudes $-17$
and $-17.5$ which has no obvious explanation. Except for this feature,
the GLFs are quite similar in both bands, and the faint end slopes are
quite flat:  $\alpha = -0.85$.

On the other hand, for Abell~1758 South (Fig.~\ref{fig:GLF_A1758S}),
the GLFs are not well fit by Schechter functions. In the $g'$ band,
there is a strong excess of galaxies at bright magnitudes, and the
faint end slope is quite flat ($\alpha = -0.59$), implying that star
formation is weak in faint galaxies of the South cluster.  In the $r'$
band, the GLF also shows an excess at very bright magnitudes, and a
strong dip for ${\rm M_{r'} \sim -17.5}$. This dip is also seen in the
$g'$ band around $-17$, though it is not as pronounced as in the $r'$
band.  Note that in the South cluster the GLFs in the $g'$ and $r'$
bands have quite different faint end slopes: $\alpha = -0.59$ in $g'$
and $\alpha = -1.11$ in $r'$. This lack of blue faint galaxies could
suggest that star formation has been quenched by a process linked to
the merger, but could also be an artefact due to the method used here
to derive the GLF (see end of Sect.~3.4).

The fact that both clusters have GLFs differing from simple Schechter
functions is most probably due to the fact that both are undergoing
merging processes, as already pointed out by \citet{DK04}
from their X-ray study.  We will discuss these results in the next
Section when considering the temperature and metallicity distributions
of the X-ray gas. These maps confirm that both clusters are indeed
structures which are strongly pertubed by several mergers, so it is
not surprising to see effects on the GLFs.

Altogether, the GLFs in Abell~1758 do not strongly differ from those
derived in other clusters.  The bright parts of the GLF Schechter fits
(${\rm M_{r'}<-19}$) for both clusters are quite similar in shape to
the GLFs recently obtained by other authors \citep[see for
example][]{andreon08}.  And the faint end slopes are within the broad
range of values estimated by previous authors for different clusters,
cluster regions and photometric bands \citep[see e.g. the compilation
in Table~A1 of][]{boue08}, except for the South cluster in the $g'$
band, where the GLF seems unusually flat.

Note that Abell~1758 is at redshift 0.279, and few GLFs are available
for clusters at such redshifts.  \citet{andreon05} found more or
less comparable faint end slopes of $-1.03$ and $-1.30$ for two
clusters at redshifts $\sim 0.3$, but in the K band, so the comparison
with our results is not straightforward.

\subsection{Galaxy luminosity functions only based on a colour-magnitude selection}
\label{sec:glf2}

As a test to our method, we also derived the GLFs by considering that
all the galaxies located within $\pm 1.5\sigma$ of the
colour-magnitude relation shown in Fig.~\ref{fig:coulmag} were cluster
members, for all magnitudes.

The corresponding points and GLF Schechter fits are shown in blue in
Figs.~\ref{fig:GLF_A1758N} and \ref{fig:GLF_A1758S} and the
corresponding Schechter parameters are given in the second half of
Table~\ref{tab:schechter}. We can see that for absolute magnitudes
fainter than about $-19$ to $-18$ the data points of the GLFs start to
differ.  The agreement between the Schechter fits based on the two
methods is fair for the North cluster in both bands and for the South
cluster in the $r'$ band, though the Schechter parameters found in
corresponding cases are not always within error bars.  This strongly
suggests that the errors on these parameters are underestimated, and
this is probably also the case for the errors on the GLF points
themselves.  On the other hand, the agreement is poor for the South
cluster in the $g'$ band. Therefore we cannot consider that the $g'$
band GLF is well constrained in the South cluster.

These results clearly illustrate the difficulty to estimate GLFs,
particularly at faint magnitudes, as already mentioned by a number of
authors (see discussion in Durret et al. 2010). They also justify our
choice not to attempt GLF fits with a higher number of free
parameters, as would be obtained by fitting a gaussian at bright
magnitudes plus a Schechter function at faint magnitudes, or two
Schechter functions.  

\subsection{Simulated galaxy luminosity function}

\begin{figure} 
\centering \mbox{\psfig{figure=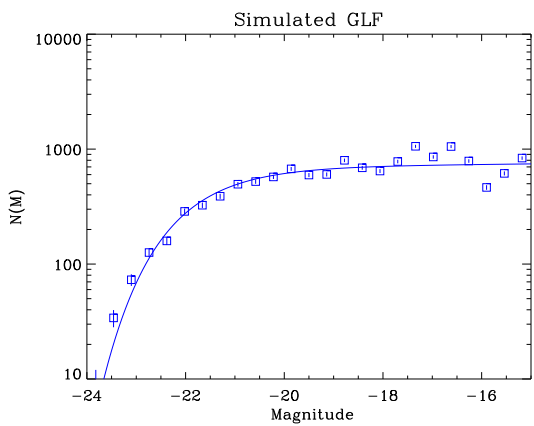,width=8cm}}
\caption{Simulated luminosity function with best Schechter fit
  superimposed.}
\label{fig:GLF_sim}
\end{figure}

In connection with the numerical simulations presented in
Section~5, we simulated a galaxy luminosity function using the halo
mass--luminosity relation from Vale \& Ostriker (2006). The halo
mass for the simulated galaxies is taken from the galaxy formation
model, which calculates the halo mass from the N-body simulation.
All the simulated galaxies were put into 25 magnitude bins between
absolute magnitudes $-25$ and $-15$.

The resulting GLF is shown in Fig.~\ref{fig:GLF_sim}. We can see
that it appears quite similar in shape to those shown in
Figs.~\ref{fig:GLF_A1758N} and \ref{fig:GLF_A1758S}. A Schechter fit
to this function is superimposed in Fig.~\ref{fig:GLF_sim} and its
parameters are given in Table~2. We can see that although the value
of $M^*$ is between half a magnitude and a magnitude brighter, the
faint-end slope agrees well with the values derived from the
observations.

\subsection{Do the North and South clusters have a cD galaxy?}

\begin{figure}[ht!]  
\centering
\includegraphics[width=0.44\textwidth,clip=true]{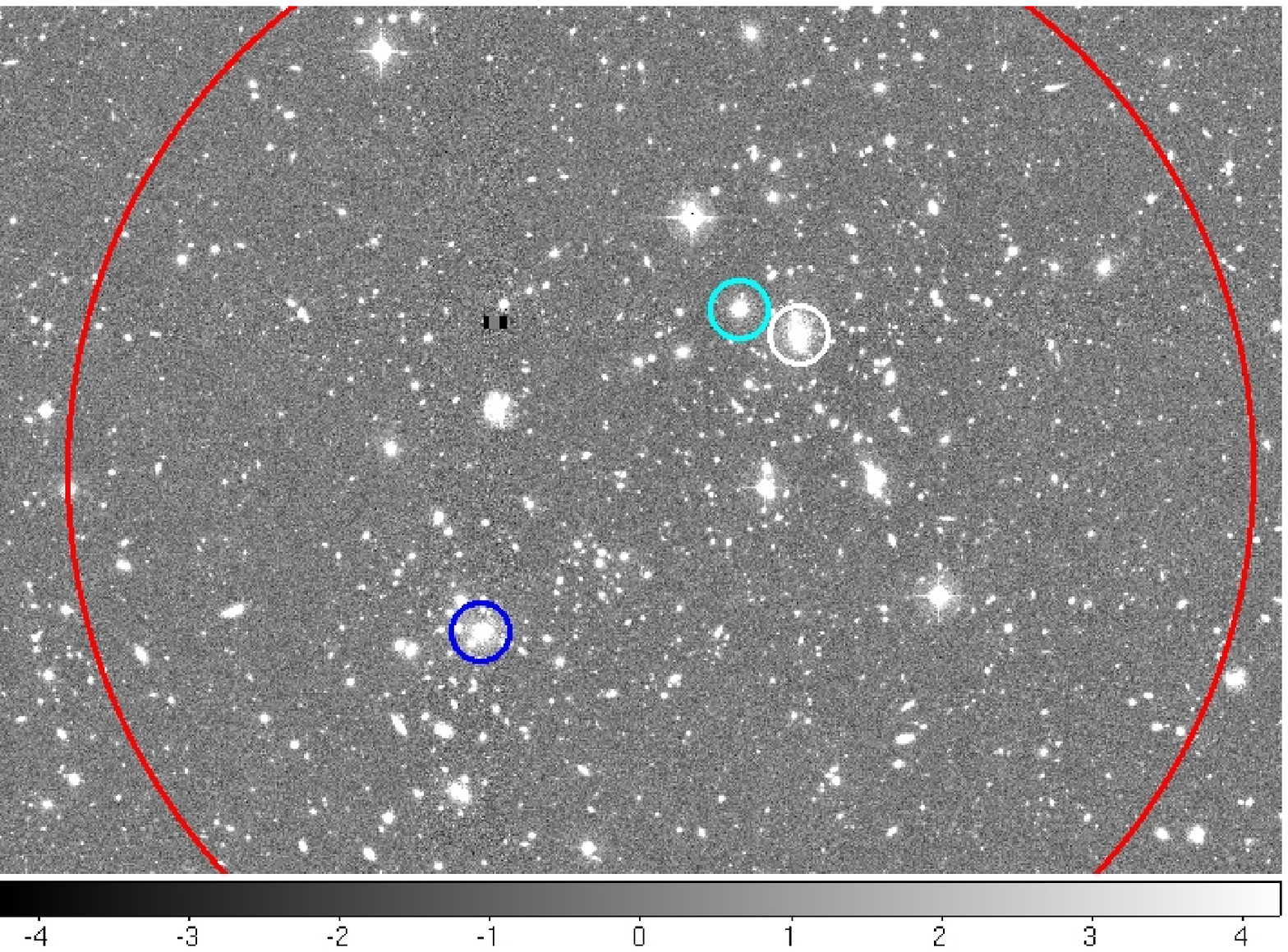}
\includegraphics[width=0.44\textwidth,clip=true]{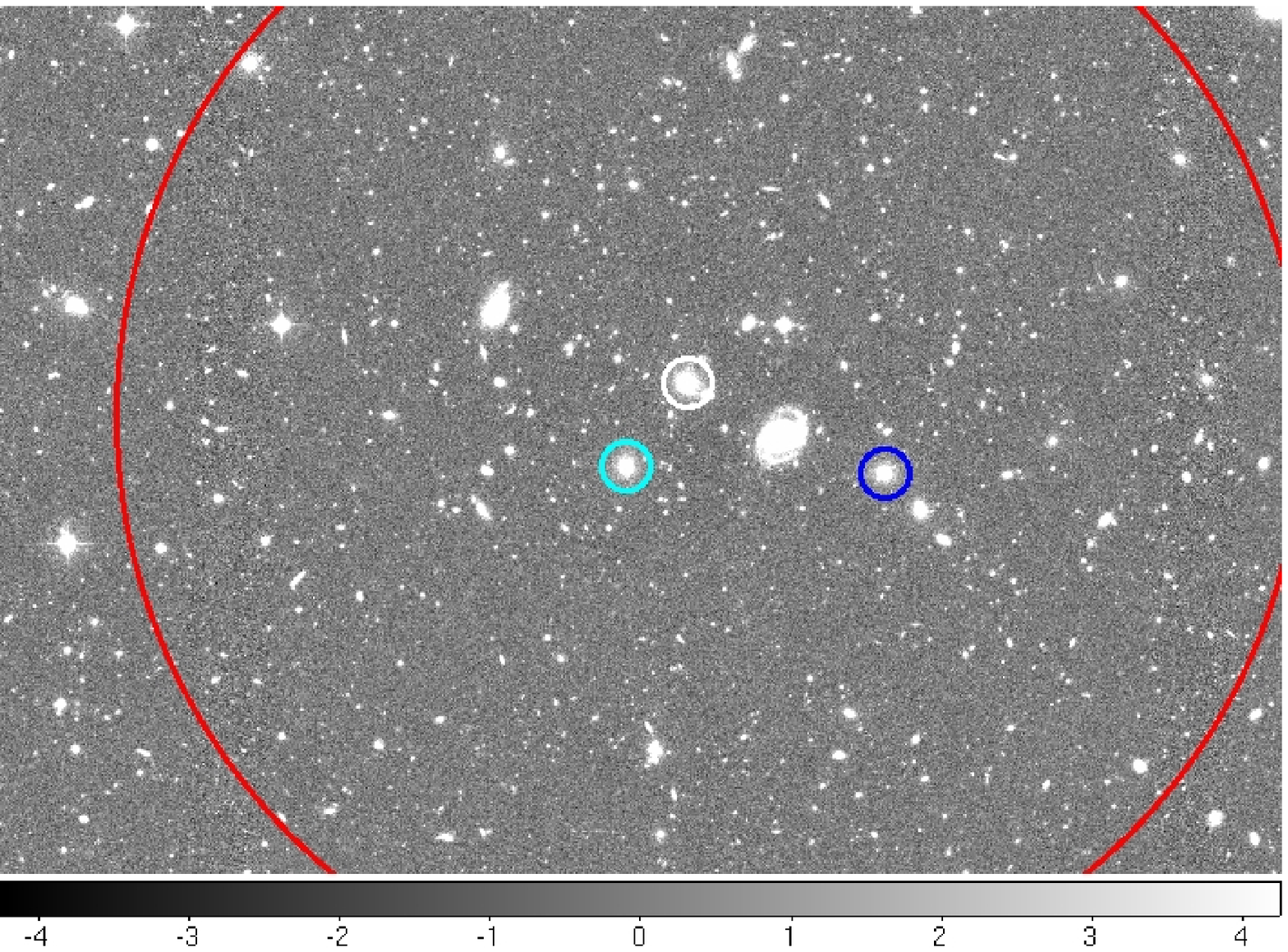}
\caption{Optical images of Abell~1758 North (top) and South
  (bottom). For each cluster, the brightest galaxy is circled in
  black, the second brightest in blue and the third brightest in
  cyan. The red circles show the cluster limits. The uncircled galaxy
  near the center of the South cluster is a foreground object.}
\label{fig:brightestgals}
\end{figure}

\begin{figure}[ht!]  
\centering
\includegraphics[width=0.4\textwidth]{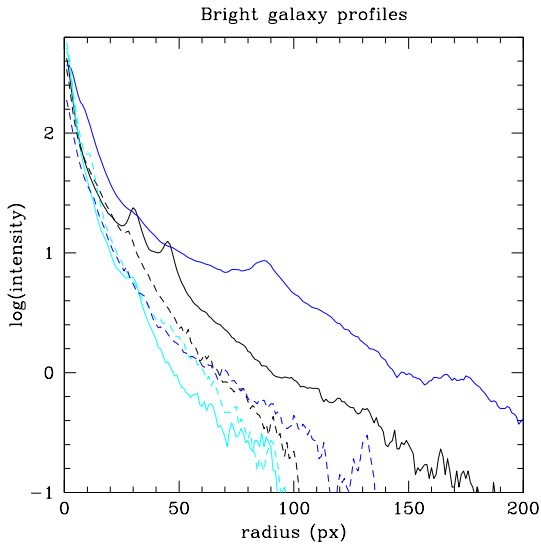}
\caption{Surface brightness profiles of the three brightest galaxies
  in both clusters (intensities are normalized to the central value
  and in arbitrary units, and radii are in pixels). Full and dashed
  lines correspond to galaxies in the North and South clusters
  respectively. For each cluster, the brightest galaxy is in black,
  the second brightest in blue and the third brightest in cyan.}
\label{fig:profils}
\end{figure}

The images of the three brightest galaxies of each cluster are shown
in Fig.~\ref{fig:brightestgals}. We can see that in the North cluster
none of these galaxies is at the cluster centre, and there are
obviously two subclusters. In the South cluster, none of the three
brightest galaxies is perfectly in the centre either (the galaxy which
is not circled is a foreground object).

The similarity with the Abell~222/223 cluster pair
\citep[see][~figure~20]{durret10} has led us to analyze the brightness
profiles of the brightest galaxies in the Abell~1758 North and South
clusters.  We had found that Abell~223 (which resembles Abell~1758
North) had two Brightest Cluster Galaxies (BCGs), one of them showing a
brightness profile decreasing slowlier than the other, and therefore
resembling that of a cD, while the other one could be the central
galaxy of an accreted group.

The surface brightness profiles of the three brightest galaxies in the
two Abell~1758 clusters are displayed in Fig.~\ref{fig:profils}. We
can see that the profiles of the two brightest galaxies of the North
cluster decrease notably slowlier with radius that those of the other
bright galaxies, and surprisingly, the profile of the second brightest
galaxy is much flatter than that of the brightest one. This suggests
that there are two dominant galaxies in the North cluster, quite
similarly to what was seen in Abell~223, and this is another
indication of a merger. The fact of having two dominant galaxies is
probably another indication of a merger of two smaller systems, in
agreement with \citet{DK04} who suggested that at least two
smaller clusters have crossed this North system.  On the other
hand, there is no dominant galaxy in the South cluster, where the
profiles of the three brightest galaxies are comparable.

\section{X-ray analysis}
\label{Xrayana}

\begin{figure}[ht!]  
\centering
\includegraphics[width=0.45\textwidth,clip=true]{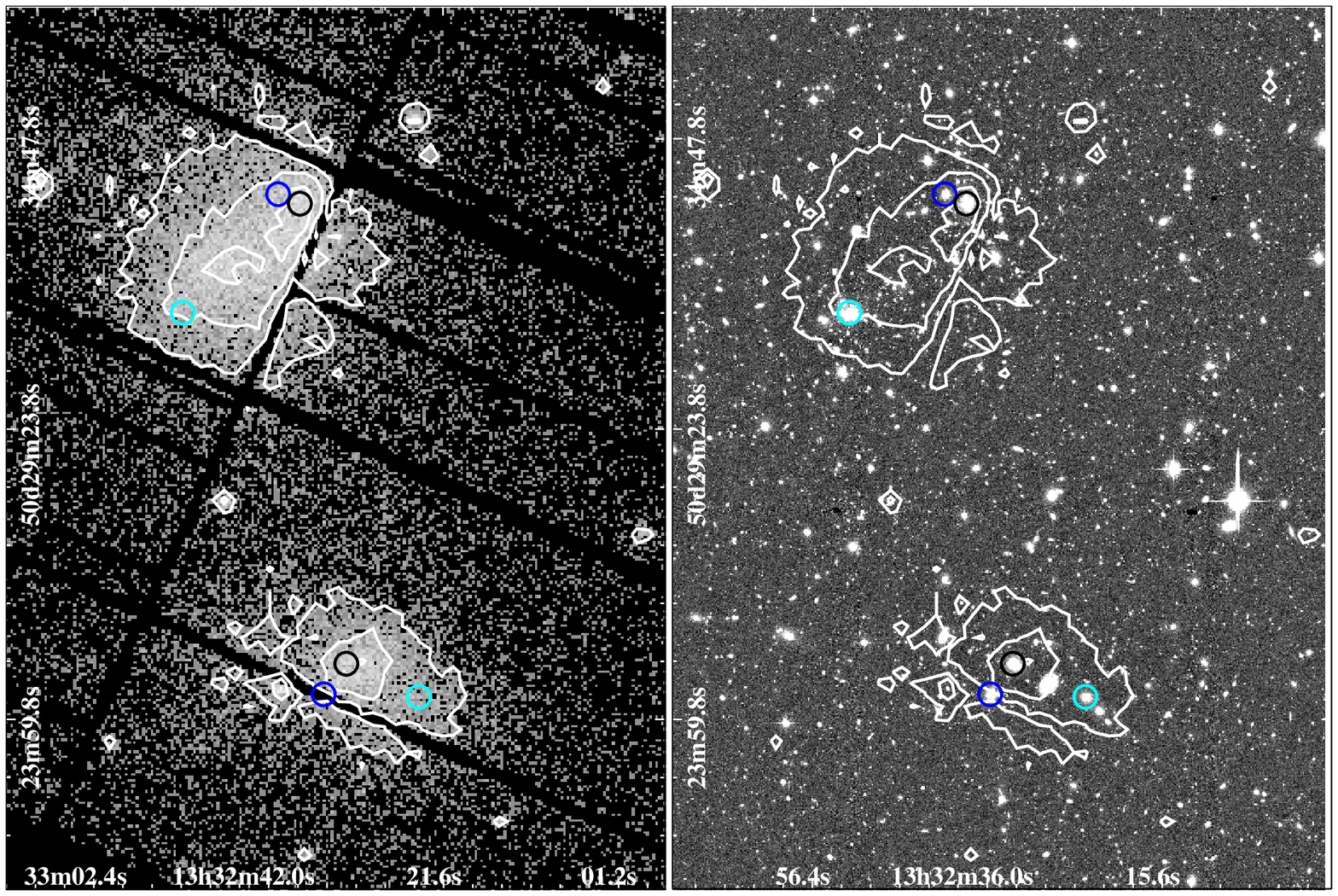}
\caption{Optical image with X-ray isocontours superimposed.  The circles
  correspond to the three brightest galaxies in Abell~1758 North.  The
  colors are the same as in Fig.~\ref{fig:profils}: the brightest
  galaxy is in black, the second brightest in blue and the third
  brightest in cyan.}
\label{SB}
\end{figure}

We present in Fig.~\ref{SB} the X-ray surface brightness isocontours
oveplotted on the optical $r'$ band image of Abell~1758 and indicate
the positions of the three brightest galaxies of each cluster.  The
three galaxies of each cluster are the remaining central galaxies from
the ancient sub-clusters. Another important thing to notice is their
position: the two brightest galaxies one near the other and the third
one is radially opposite.  This is an indication of a recent merger
where we clearly see that while the gas has already settled down, the
brightest galaxies are not in the center of Abell~1758 North and
A1758~South. We will revisit this merging scenario in the following
sections.

\subsection{Data reduction}
Abell~1758 was observed for $\sim$57~ks with XMM-\textit{Newton} with
the Medium filter inserted. The XMM-\textit{Newton} ODF files were
processed using SAS version v8.0.  The MOS and pn files were filtered
excluding all events with FLAG $>$ 0 and PATTERN $>$ 12, and FLAG $>$ 0
and PATTERN $>$ 4, respectively.  Light curves were made in the
 1--10  keV energy band and periods where the background value exceeded
the mean value by more than 3$\sigma$ were excluded. We considered
events inside the field of view (FOV) and excluded all bad pixels.
Flare filtering left live times of 16928~s, 12734~s and 16929~s in the
MOS1, MOS2 and pn cameras respectively.

The background was taken into account by extracting MOS1,
MOS2 and pn spectra from the publicly available EPIC blank
sky templates of Andy Read (Read \& Ponman 2003). The background
was normalized using a spectrum obtained in an annulus
(between 12.5--14 arcmin) where the cluster emission is no
longer detected.

\subsection{Spectrally measured 2D X-ray maps} 

We performed quantitative studies using X-ray spectrally measured 2D
maps to derive global properties of these two clusters.  These maps
were made in a grid; for each spatial bin we set a minimum count
number of 900 (after background subtraction).  For the spectral fits,
we used XSPEC version 11.0.1 \citep{arnaud96} and modeled the obtained
spectra with a MEKAL single temperature plasma emission model 
\citep[bremsstrahlung + line emission][]{kaastra93,liedahl95}.  
The free parameters are the X-ray temperature (kT) and the
metal abundance (metallicity). Spectral fits were made in the energy
interval of  0.7$-$8.0   keV with the hydrogen column density fixed at
the Galactic value ($1.06 \times 10^{20}~\rm cm^{-2}$), estimated with
the nH task of FTOOLS \citep[based on][]{DL90}.

We compute the effective area files (ARFs) and the response matrices
(RMFs) for each region in the grid.  This procedure \citep[already described
in][]{durret10} allows us to perform a reliable spectral
analysis in each spatial bin, in order to derive high precision
temperature and metallicity maps, since we simultaneously fit all
three instruments. The best fit value is then attributed to the
central pixel.

\begin{figure}[ht!]  
\centering
\includegraphics[width=0.45\textwidth]{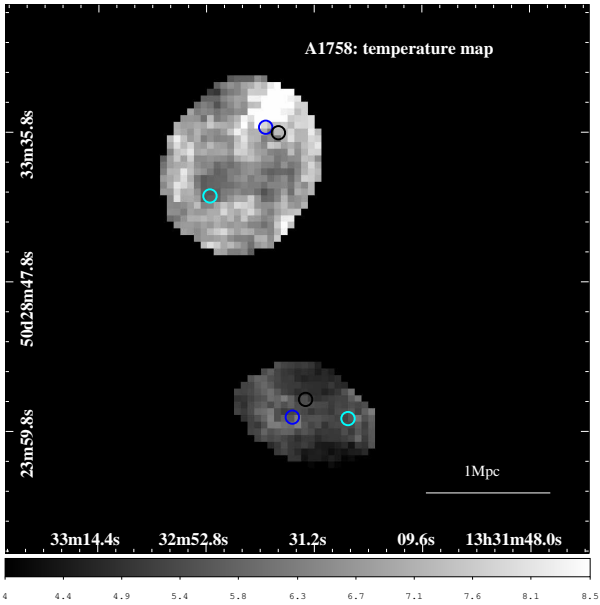}
\caption{X-ray temperature map for Abell~1758 North (top) and South
  (bottom).  The color-bar indicates the temperature in keV. The
  circles correspond to the three brightest galaxies in each
  cluster. The colors are the same as in Fig.~\ref{fig:profils}: the
  brightest galaxy is in black, the second brightest in blue and the
  third brightest in cyan.  The corresponding error map is shown in
  Fig.~\ref{err_maps} (top).}
\label{map_kT}
\end{figure}

\begin{figure} ht!  
\centering
\includegraphics[width=0.45\textwidth]{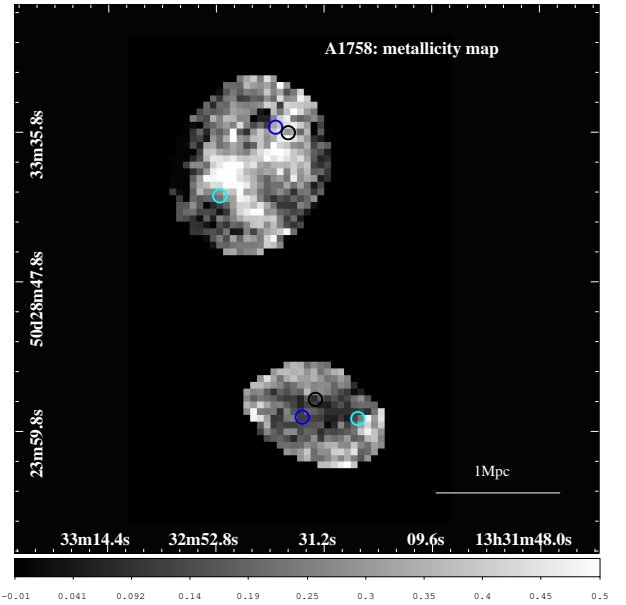}
\caption{X-ray metallicity map for Abell~1758 North (top) and South
  (bottom).  The color-bar indicates the metallicity in solar units.
  The circles correspond to the three brightest galaxies in each
  cluster. The colors are the same as in Fig.~\ref{fig:profils}: the
  brightest galaxy is in black, the second brightest in blue and the
  third brightest in cyan.  The corresponding error map is shown in
  Fig.~\ref{err_maps} (bottom).}
\label{map_Z}
\end{figure}

The resulting X-ray temperature and metallicity maps are displayed in
Figs.~\ref{map_kT} and \ref{map_Z}. Maps of the errors on these two
maps are given in the Appendix (Fig.~\ref{err_maps}). 

The temperature maps in Fig.~\ref{map_kT} show that both clusters
appear almost isothermal, the North one being hotter ($\sim 6-7$~keV)
and the South one being cooler ($\sim 4-5$~keV).  The presence of
inhomogeneities is not obvious in the temperature maps, except for a
hotter blob in the northwest region of Abell~1758 North.  Although
signs of recent interactions are not clearly present in the
temperature map, the positions of the three brightest galaxies argue
in favour of a merging scenario where two clusters merged to form
Abell~1758 North, as mentioned in Sect.~\ref{Xrayana}.

On the other hand, two striking aspects call the attention when
considering the metallicity map shown in Fig.~\ref{map_Z}. First, we
can note two elongated regions of high metallicity in the North
cluster, suggesting that at least two smaller clusters have crossed
the North cluster \citep[as pointed out by][]{DK04}. Based on the
  positions of the three brightest galaxies we can assume a scenario in
  which these elongated regions are due to metals were ton by
  ram-pressure stripping during the merger.

Second, the metallicity map of the South cluster is even more unusual,
since it shows a deficit of metals in the central region. This deficit
is probably the signature of an interaction with the central
object. Since this cluster is less massive, the effects of galactic
winds or supernova explosions will be stronger to expell metals
towards the outskirts \citep{evrard08}.  

However, none of the metal structures identified in the 2D maps
correlates with any temperature structure for either cluster. We do
not see the high-metallicity regions of the North cluster in the
temperature map.

To have an overview of the merging scenario and better understand the
dynamical history of these clusters, we considered the results of
numerical simulations to give support to our findings, and compared
them with our observational results. This will be presented in the
next section.

\citet{DK04} have already analyzed this system using
XMM-\textit{Newton} and \textit{Chandra} data. They conclude that
these two clusters most likely form a gravitationally bound system,
though their imaging and spectroscopic analyses do not reveal any sign
of interaction between the North and South clusters, and show that
A1758N and A1758S are both undergoing major mergers.  It is clear from
our analysis that Abell~1758 North is still undergoing at least one
merger, the bright and hot zone seen in the temperature map probably
being the signature of one of the cores. For Abell~1758 South, the
merger is not evident in X-rays, though some kind of interaction is
suggested by the absence of a cool core, and confirmed by the optical
data analysis.

\section{Numerical simulation results}

As described in detail in \citet{kapferer06} \citet{kapferer07} or
\citet{schindler05} we combine several codes to model the metal
distribution in a galaxy cluster. To calculate the dark matter
structure of the galaxy cluster we use the N-body code GADGET2
\citep{springel05} with constrained random fields as initial
conditions \citep{hoffman91}.A semi-analytical galaxy-formation model
then assigns galaxy properties to the halos found in the dark matter
structure. For this galaxy-formation model we use an improved version
of the code described by \citet{kampen99}. The chemical evolution of
the galaxies is modeled as in \citet{matteucci89}. To study the
evolution of the ICM we use a hydrodynamic grid code with comoving
coordinates and a PPM-scheme for a better treatment of shocks
\citep{colella84}.  To optimize the computational time, we use 4
nested grids \citep{ruffert92}, each with 128x128x128 cells. The total
computational hydro-volume has a side length of 20 Mpc. We included
special routines to transport enriched material out from the modelled
galaxies into the ICM: we model ram-pressure stripping as well as
galactic winds as described in
\citet{GG72,domainko06,kapferer07}. More specifically, the
ram-pressure and galactic wind algorithms calculate a mass loss for
each model galaxy (depending on its velocity relative to the ICM and
its star formation rate). As the galaxy formation model provides us
with the metallicity within the ISM, we know the quantity of metals
transported from the ISM into the ICM. These metals are then advected
with the ICM, and by knowing the ICM density we can then calculate the
ICM metallicity. Combining these codes, we can simulate the 3D
evolution of the ICM and its metallicity. From these 3D data, we then
extract temperature and metal maps, by making emission weighted
projections. These maps can be compared directly to observations.

\begin{figure*}[ht!]  
\centering
\includegraphics[width=0.45\textwidth]{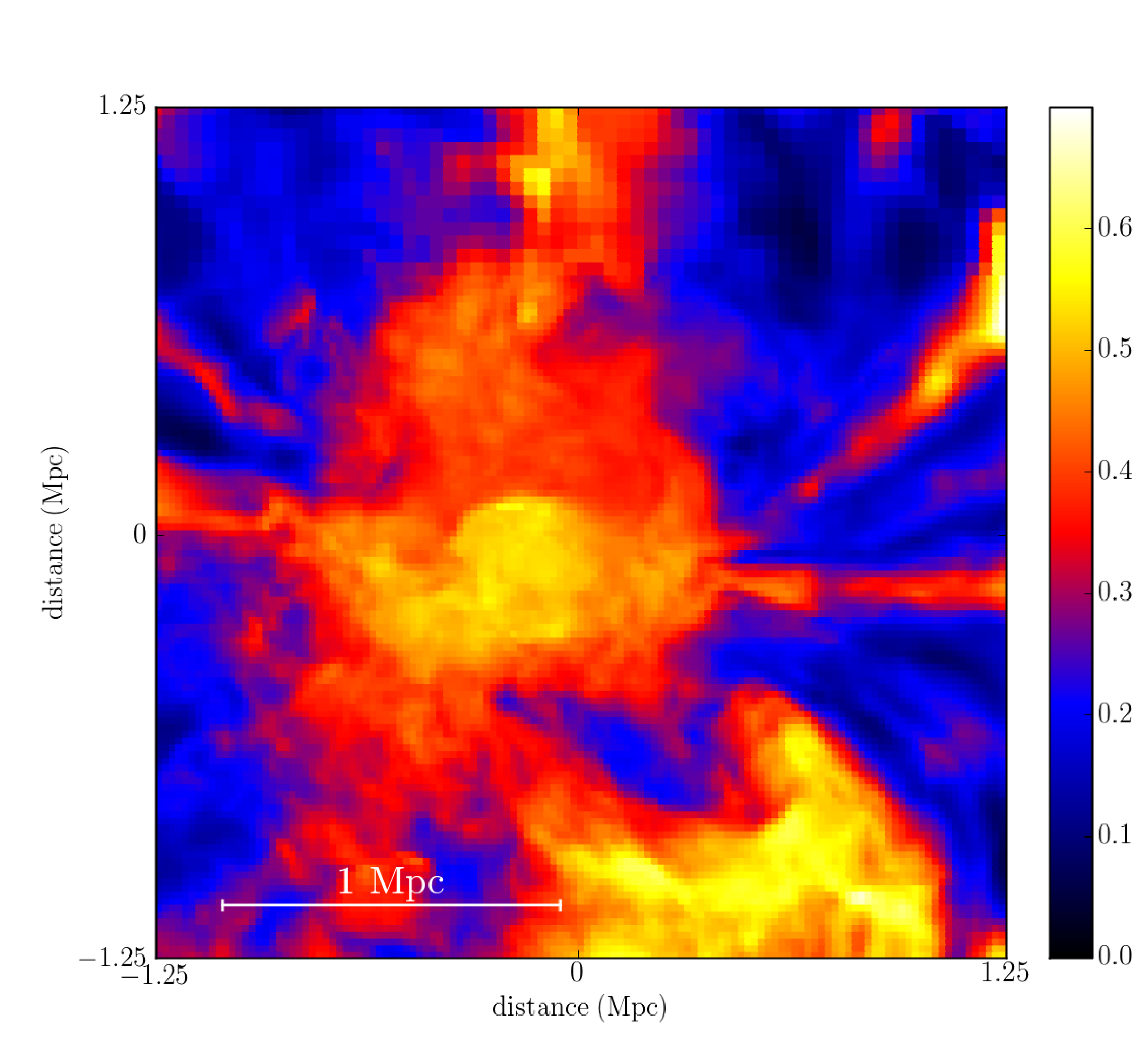}
\includegraphics[width=0.40\textwidth]{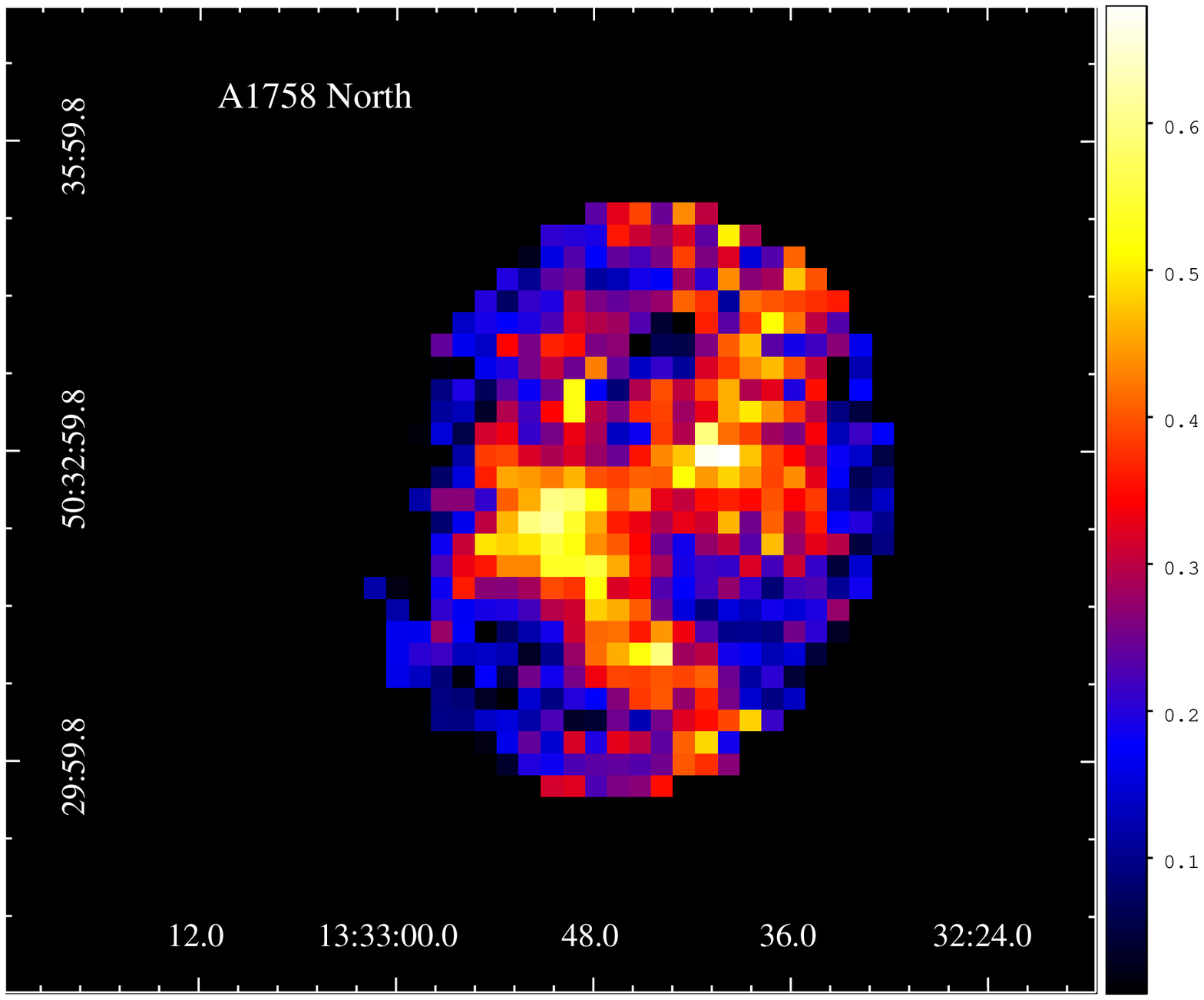}
\includegraphics[width=0.9\textwidth,clip=true]{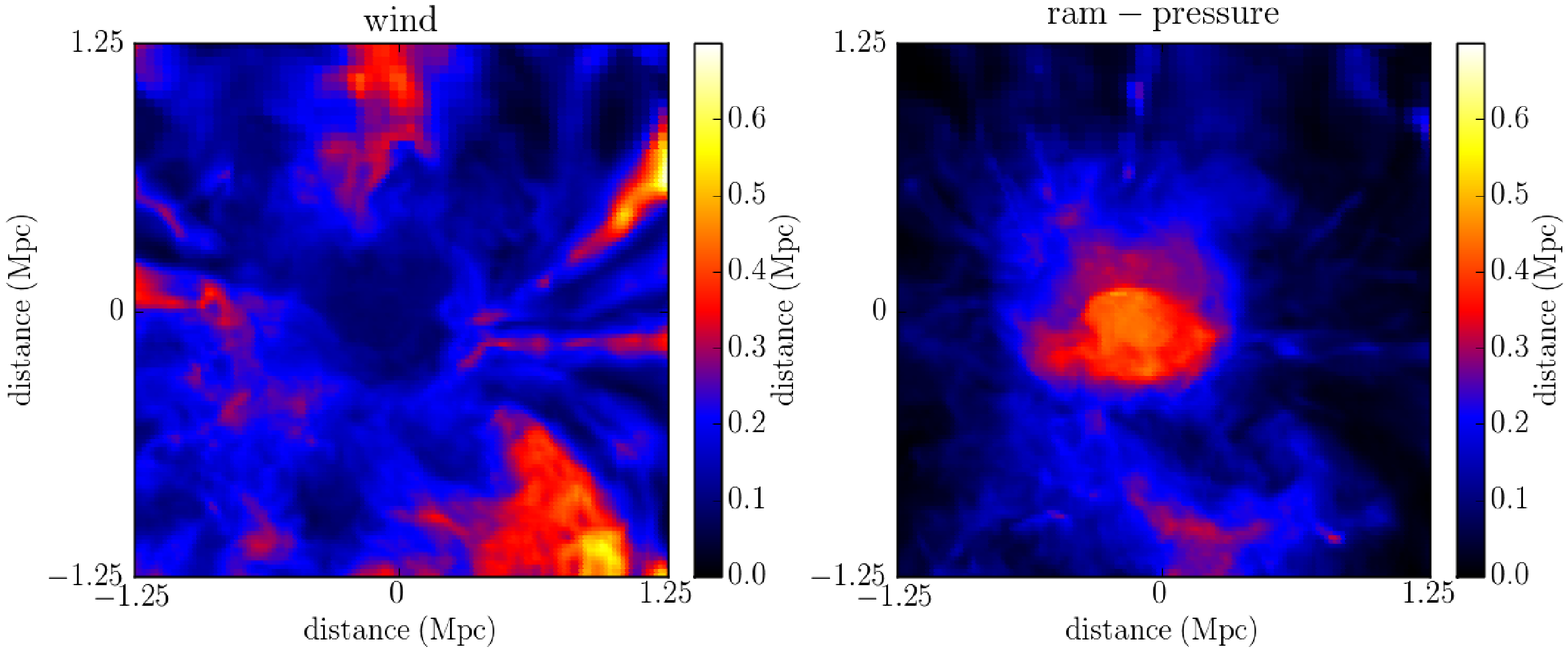}
\caption{\textit{Top left panel}: metal distribution predicted by
  numerical simulations at the cluster redshift (i.e. sum of the two
  panels below).  \textit{Top right panel}: observed metal
  distribution map in Abell~1758 North.  Both panels have a side-length of
  2.5 Mpc. \textit{Bottom left panel}: metals which were ejected by
  galactic winds. \textit{Bottom right panel}: metals which were
  transported out of the galaxies by ram-pressure stripping.}
\label{fig:mapa_z_sim_zoom}
\end{figure*}


We performed 5 simulations with different initial conditions and
selected the simulation which best matched our observational
results. It should be noted that we did not set up the simulations to
specifically reproduce Abell~1758.
Among all the tests perfomed here, we found one metal distribution
that reproduces quite reasonably the elongated region of high abundance
found observationally for the North cluster. We see that the
simulated 2D map in Fig.~\ref{fig:mapa_z_sim_zoom} can also reproduce the
exact values for the metallicity of the North cluster. For the bright
regions (in yellow color) we have metallicity values of about
$0.5-0.6$ solar units.  The metallicity decreases towards the
outskirts where it reaches values around 0.2 solar units.

It is of great importance to mention the spatial scales of the
features presented in the 2D metal distribution maps predicted by
numerical simulations and derived from observations.  The physical
sizes of the two top panels in Fig.~\ref{fig:mapa_z_sim_zoom} are exactly the
same (a side-length of 2.5 Mpc), but the features, although similar in
shape, do not have the same physical sizes. In the observational metal
distribution (top right panel in Fig.~\ref{fig:mapa_z_sim_zoom}) we see
that Abell~1758 North is completely comprised within 2.5~Mpc, while in
the metal distribution computed from numerical simulations (top left panel
in Fig.~\ref{fig:mapa_z_sim_zoom}) the cluster extends over a larger
region.  The point here is that we did not make the simulations to
exactly match this cluster. 

In order to analyze the importance of galactic winds and ram-pressure
stripping in transporting metals, we display in
Fig.~\ref{fig:mapa_z_sim_zoom} the results of numerical simulations,
showing separately the metals ejected by galactic winds (bottom left),
those transported out of the galaxies by ram-pressure stripping
(bottom right), and the sum of the metals accounted for by these two
processes (top left). These maps can be compared to the metallicity
map derived from observations for the North cluster (top right). From
these figures, we can say that both processes are important for metal
enrichment, playing different roles in the cluster, with winds
definitely playing a higher part in the south-west corner. Galactic
winds are more important in the outskirts of the cluster, while
ram-pressure stripping seems to be dominant in the inner parts.

It is worth noting the time scales of these different enrichment
processes. By analyzing the results of our numerical simulation, we can
say that a large part of the metals for the high metallicity region in
the south of Abell~1758 North and the thin elongated stripe from the
center to the north-west have been transported into the ICM at
redshifts z$ > 3$. Then, between redshifts 3 and 2 there was a
substantial contribution to the metallicity in the elongated structure
but no contribution to the center of the cluster.  Between redshifts 2
and 1, there was a substantial contribution to the central region
(around 1/3 of the final metallicity), mainly due to
ram-pressure. Then, between redshifts 1 and 0.8 there was hardly any
contribution to the metallicity at all.  From z=0.8 to z=0.5 there was
a contribution (around 0.2 solar metallicities) to the metals in the
central region, coming only from ram-pressure stripping. And finally,
from z=0.5 to z=0.279 (the cluster redshift) there was hardly any
significant contribution to the metallicity.  In summary, many of the
metals which are responsible for the features in the outskirts have
been transported into the ICM at very early times, even before z=3. If
we look at the metals which have been transported into the ICM before
z=2, we see that they explain the main morphological features in the
outskirts (like the elongated structure). On the other hand, the
enrichment of the central region takes place mainly between z=2 and
z=0.5, predominantly via ram-pressure stripping.

The results presented above show that the comparison of observations
with the results of numerical simulations is a powerful tool to
understand better the physical processes involved in the transport of
metals. However, there is still work to be done on this topic, based
on a larger sample of clusters in different dynamical states.

\section{Discussion and Conclusions}

It is becoming clear that clusters are in a continual state of
evolution and with high precision telescopes, such as
XMM-\textit{Newton} and \textit{Chandra}, we see that they are hardly
ever relaxed and virialized as it was previously assumed.  We have
thus become interested in multiple mergers, where merging effects are
expected to be even stronger.

As a second study of cluster pairs, we chose to analyse Abell~1758 (at
a redshift of 0.279) by coupling archive optical CFHT Megacam and
XMM-\textit{Newton} data, and by comparing the temperature and
metallicy maps obtained for the ICM with the results of numerical
simulations. We made a step further by also considering the results of
numerical simulations to try to understand better the dynamics and
building-up history of this system.

From our results, signs of merger(s) are detected in the optical as
well as X-ray wavelength ranges, meaning that both galaxies and gas
are still out of equilibrium.  From optical results we see that for
Abell~1758 North, a Schechter function fits rather well most of the
points of the GLFs in the $g'$ band, but there is an excess of
galaxies over a Schechter function in the brightest magnitude bins,
specially in the $r'$ band, as well as a possible and unexplained
excess of galaxies around M$_{r'}\sim -17.5$. On the other hand, for
Abell~1758 South, which is poorer than the North cluster, the GLF is
not as well defined and fit by a Schechter function. Note in
particular that the GLFs derived by the two methods (statistical
background subtraction for $r'>22$ or colour-magnitude selection at
all magnitudes)) look quite different, particularly in the $g'$
band. If the dip detected in the $r'$ band GLF for M${_{r'} \sim -18}$
is real, as already found in other clusters but at somewhat brighter
absolute magnitudes, it could be interpreted as showing the transition
zone between (brighter) ellipticals and (fainter) dwarfs
\citep{durret99}. An excess of galaxies over a Schechter function in
the brightest magnitude bins is observed in both bands.  We can also
note that in the South cluster the GLFs in the $g'$ and $r'$ bands
look quite different, with a faint end slope somewhat steeper in $r'$
than in $g'$. The shallower faint end slope in $g'$, if real, could be
explained either by quenching of star formation due to the merger
(which strips galaxies from their gas and reduces star formation), or
by the fact that star formation has not yet had time to be triggered
by the merger \citep[see e.g.][]{bekki99}.  The somewhat perturbed
shapes of the GLFs of both clusters therefore agree with the fact that
they are both undergoing merging processes.

From the X-ray analysis we have noticed that the gas temperature maps
do not present prominent inhomogeneities, except for a hotter blob in
the northwest of Abell~1758 North.  The North cluster is hotter (with
temperatures in the range of $\sim 6-7$~keV) and the South one cooler
(with kT $\sim 4-5$~keV). The hotter blob in the northwest of
Abell~1758 North could be explained by heating of the gas in that
region by the movement of the northwest system towards the north
proposed by \citet{DK04}.  The most striking features are
seen in the metallicity maps. The metallicity map of the South cluster
is even more unusual, since it shows a deficit of metals in the
central region. This deficit is probably the signature of an
interaction with the central object that could have expelled metals
towards the outskirts \citep[see][and top left panel of 
Fig.~\ref{fig:mapa_z_sim_zoom}]{evrard08}.  
We also detect two elongated
regions of high metallicity in the North cluster, suggesting that at
least two smaller clusters have crossed the North cluster.  Note that
our temperature and metallicity maps are in agreement with the
scenarios proposed by \citet{DK04}, who derived that the
North cluster is in the later stages of a large impact parameter
merger between two 7~keV clusters, while the south cluster is in the
earlier stages of a nearly head--on merger between two 5~keV clusters.


In order to understand better the nature of the most prominent
features exhibited in the metallicity map of the North cluster, we
performed 5 simulations with different initial conditions.  It should
be stressed that we did not set up the simulations to specifically
reproduce Abell~1758.  Among these 5 simulations, we found one metal
distribution that reproduces quite reasonably the elongated region of
high abundance found observationally for the North cluster, although
without a perfect spatial correlation. The results of our numerical
simulations allowed us to distinguish the role of metal transportation
processes such as galactic winds and ram-pressure stripping. We have
shown that these phenomena act in different regions of the cluster,
and that in the metal-rich elongated regions of the North cluster,
winds were more efficient in transporting enhanced gas to the
outskirts than ram pressure stripping. These simulations also allowed
us to compute the GLF, and the result is consistent with the observed
GLFs.

In the hope of finding some kind of large scale structure and/or
filaments linking Abell~1758 with its surroundings, we searched the
NED database for galaxies with redshifts available in a region of
36~arcmin around Abell~1758 (corresponding to 9.1~Mpc at the cluster
distance). We found 137 galaxies, out of which only 15 have redshifts
in the approximate cluster redshift range:  0.264, 0.294  . No
structure is found in the blueshifted and redshifted galaxies, except
for what appears to be a cluster west of Abell~1758 (see Section 3.2).
Therefore we cannot derive any argument from the large scale
distribution of galaxies around Abell~1758.

To conclude, the Abell~1758 North and South clusters most likely form
a gravitationally bound system, but our imaging analyses of the X-ray
and optical data do not reveal any sign of interaction between the two
clusters.  All signs of dynamical disturbance are associated with
recent merger(s) which each of the two clusters is still undergoing.

\begin{acknowledgements}

  We are grateful to Christophe Adami, Gast\~ao B. Lima Neto and
  Sabine Schindler for discussions. We warmly thank Andrea Biviano for
  giving us his Schechter function fitting programme and helping us
  with the corresponding plots. Thanks also to Thad Szabo for sending
  us information prior to publication. FD acknowledges long-term
  support from CNES. TFL thanks financial support from FAPESP (grants:
  2006/56213-9, 2008/04318-7).  MH is grateful to the Austrian Science
  Foundation (FWF) through grant number P19300 and for support by the
  Austrian Ministry of Science BMWF as part of the
  UniInfrastrukturprogramm of the Forschungsplattform Scientific
  Computing at the LFU Innsbruck. Finally, we acknowledge the
  referee's interesting and constructive comments.

\end{acknowledgements}

\appendix

\section{Error maps}
\label{err}
\begin{figure}[ht!]  
\centering
\includegraphics[width=0.35\textwidth]{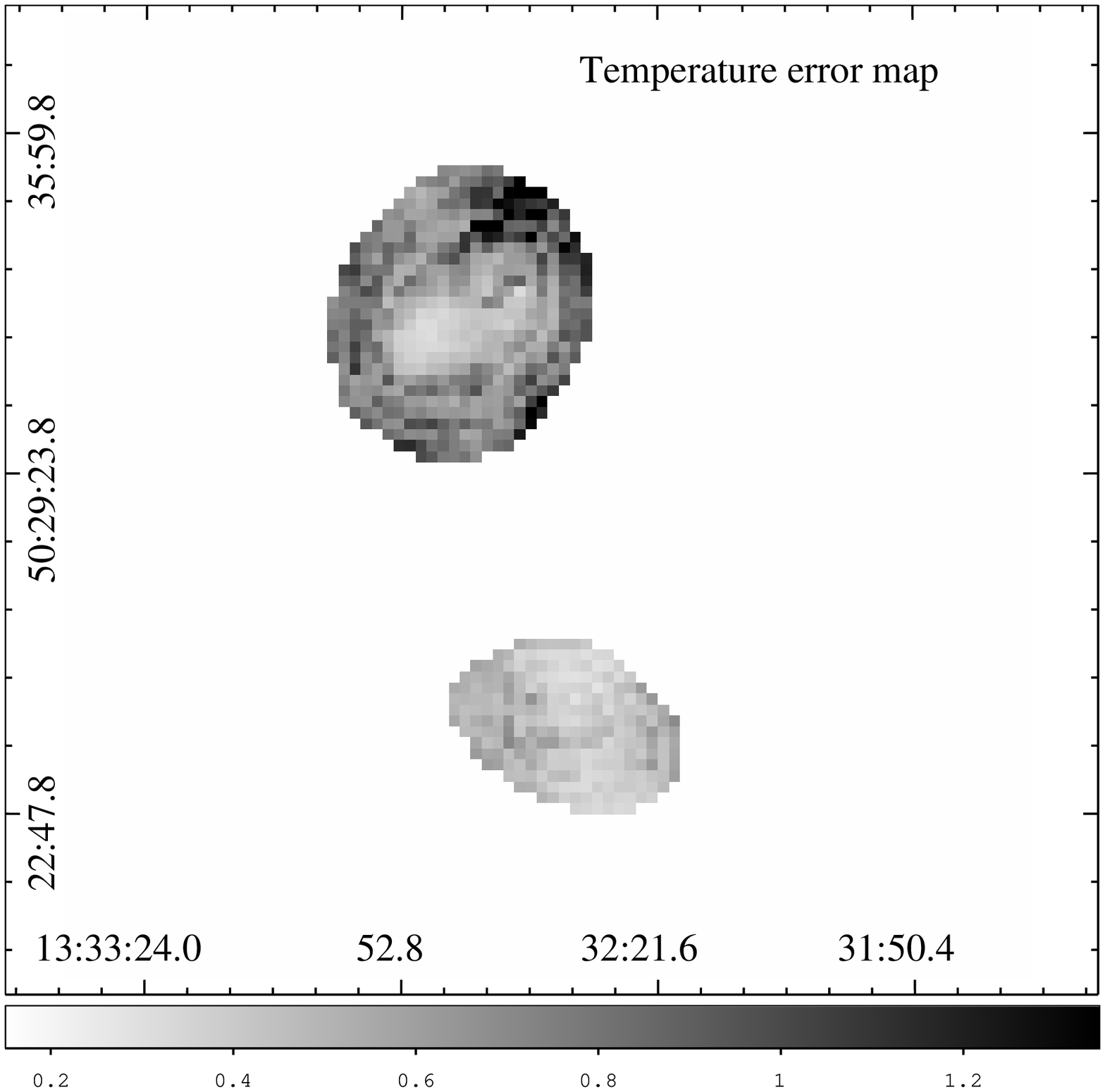}
\includegraphics[width=0.35\textwidth]{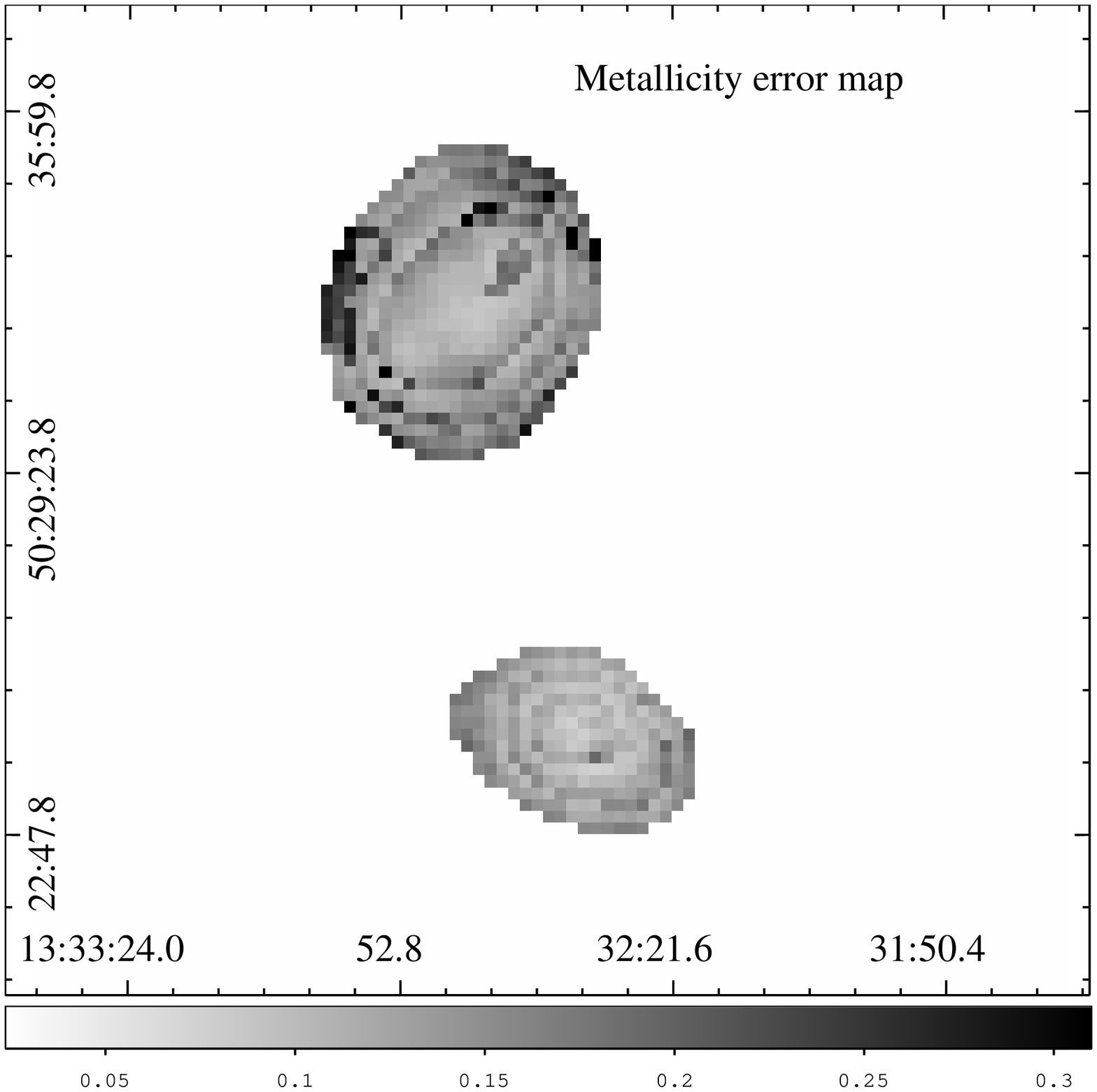}
\caption{\textit{Upper panel}: error map on the X-ray temperature.
\textit{Lower panel}: error map on the X-ray metallicity.}
\label{err_maps}
\end{figure}

In Fig.~\ref{err_maps} we present the errors associated to each bin in
the temperature and metallicity maps displayed in Figs.~\ref{map_kT}
and \ref{map_Z}.  The errors on these parameters were directly
obtained from the spectral fits.

Looking at Figs.~\ref{map_kT}, \ref{map_Z} and \ref{err_maps} (error
maps) we see that, for the temperature estimates, we have errors
around 5\% for A1758 South, while for A1758 North they vary from 5\%
(in the inner parts) up to 12\% (at the outskirts). The errors on the
temperature maps are therefore very reasonable. For the metallicity
maps we do not have the same accuracy.  Looking at the same figures,
we see that for A1758 South, the metallicity errors can reach 25\% and
for A1758 North they vary from 30\% (in the inner region) up to 46\%
(at the outskirts). It is important to notice that although the
XMM-Newton exposure time on Abell~1758 was about 57~ks, due to flare
filtering only $\approx$ 17~ks of good data were useable to construct
the 2D spectral maps.


\begin{thebibliography}{}

\bibitem[Abazajian et al. (2009)]{Abazajian09} Abazajian K. N., Adelman-McCarthy J. K., 
Ag\"ueros M. A. et al. 2009, ApJS, 182, 543.

\bibitem[Adami et al. (1998)]{adami98} Adami C., Biviano A., Mazure A. 1998, A\&A 331, 439

\bibitem[Adami et al. (2006)]{adami06} Adami C., Picat J.-P., Savine C. et al. 2006, A\&A 451, 1159 

\bibitem[Adami et al. (2007)]{adami07} Adami C., Durret F., Mazure A. et al. 2007, A\&A 462, 411

\bibitem[Andreon et al. (2005)]{andreon05} Andreon S., Punzi G., Grado A. 2005, MNRAS 360, 727

\bibitem[Andreon et al. (2008)]{andreon08} Andreon S., Puddu E., de Propris R., Cuillandre J.-C. 2008,
MNRAS 385, 979

\bibitem[Arnaud (1996)]{arnaud96} Arnaud K.~A. 1996, ASPC 101, 17

\bibitem[Bekki (1999)]{bekki99} Bekki K. 1999, ApJ 510, L15

\bibitem[Bertin \& Arnouts (1996)]{bertin96} Bertin E. \& Arnouts S. 1996, A\&AS 117, 393

\bibitem[Bou\'e et al. (2008)]{boue08} Bou\'e G., Adami C., Durret F., Mamon G., Cayatte V., 2008, A\&A 479, 335

\bibitem[Bourdin et al. (2004)]{bourdin04} Bourdin H., Sauvageot J.-L., Slezak E., Bijaoui A., Teyssier R.
2004, A\&A 414, 429

\bibitem[Colella (1984)]{colella84} Colella P., Woodward P. 1984, Journal of Computational Physics, 54, 174

\bibitem[David \& Kempner (2004)]{DK04} David L.P. \& Kempner J. 2004, ApJ 613, 831  

\bibitem[Dickey \& Lockman (1990)]{DL90} Dickey J.~M. \& Lockman F.J. 1990, ARA\&A 28, 215D  

\bibitem[Dietrich, Clowe \& Soucail (2002)]{dietrich02} Dietrich J.P., Clowe D.I., Soucail G. 2002, A\&A 394, 395  

\bibitem[Dietrich et al. (2005)]{dietrich05} Dietrich J.P., Schneider P., Clowe D., Romano-D\'\i az E.,
  Kerp J. 2005, A\&A 440, 453  

\bibitem[Domainko \& Kapferer (2006)]{domainko06} Domainko W., Mair M., Kapferer W., et al. 2006, A\&A 452, 795  

\bibitem[Durret et al. (1999)]{durret99} Durret F., Gerbal D., Lobo C., Pichon C. 1999, A\&A 343, 760  

\bibitem[Durret et al. (2003)]{durret03} Durret F., Lima Neto G.~B., Forman W., Churazov E. 2003, 
A\&A 403, L29  

\bibitem[Durret, Lima Neto \& Forman (2005)]{durret05} Durret F., Lima Neto G.~B., Forman W. 2005, A\&A 432, 809  

\bibitem[Durret \& Lima Neto (2008)]{durret08} Durret F., Lima Neto G.B. 2008, AdSpR 42, 578

\bibitem[Durret et al. (2010)]{durret10} Durret F., Lagan\'a T. F., Adami C., Bertin E. 2010, A\&A 517, 94  

\bibitem[Evrard et al. (2008)]{evrard08} Evrard A. E., Bialek J., Busha M. et al. 2008, ApJ 672, 122.  

\bibitem[Gunn \& Gott (1972)]{GG72} Gunn J., Gott J. 1972, ApJ 176,1  

\bibitem[Gwyn (2009)]{gwyn09} Gwyn S.~D.~J. 2009, arXiv:0904.2568  

\bibitem[Hoffman \& Ribak (1991)]{hoffman91} Hoffman Y., Ribak E. 1991, ApJ 380  

\bibitem[Ilbert, Tresse, \& Zucca (2005)]{ilbert05} Ilbert O., Tresse L., Zucca E. et al. 2005, A\&A 439, 863  

\bibitem[Kaastra \& Mewe (1993)]{kaastra93} Kaastra J.~S., Mewe R. 1993, A\&AS 97, 443  

\bibitem[Kapferer et al. (2006)]{kapferer06} Kapferer W., Ferrari C., Domainko W., et al. 2006, A\&A 447, 827

\bibitem[Kapferer et al. (2007)]{kapferer07} Kapferer W., Kronberger T., Weratschnig J., et al. 2007, A\&A 466, 813  

\bibitem[Liedahl et al. (1995)]{liedahl95} Liedahl D.A., Osterheld A.~L., Goldstein W.~H. 1995, ApJ 438, L115  

\bibitem[Matteucci \& Fran\c cois (1989)]{matteucci89} Matteucci F., Fran\c cois P. 1989, MNRAS 239, 885  

\bibitem[Read \& Ponman (2003)]{read03} Read A. M. \& Ponman T. J., 2003, A\&A 409, 395.  

\bibitem[Robin et al. (2003)]{robin03} Robin A. C., Reyl\'e C., Derri\`ere S., Picaud S. 2003,
A\&A 409, 523  

\bibitem[Ruffert (1992)]{ruffert92} Ruffert M. 1992, A\&A 265, 82  

\bibitem[Schindler et al. (2005)]{schindler05} Schindler S., Kapferer W., Domainko W., et al. 2005, A\&A 435, L25

\bibitem[Schlegel, Finkbeiner, \& Davis (1998)]{schlegel98} Schlegel D.J., Finkbeiner D.~P., Davis M. 1998, ApJ 500, 525  

\bibitem[Springel (2005)]{springel05} Springel V. 2005, MNRAS 364, 1105

\bibitem[Vale \& Ostriker (2006)]{vale06} Vale A. \& Ostriker J.P. 2006, MNRAS 371, 1179  

\bibitem[van Kampen et al. (1999)]{kampen99} van Kampen E., Jimenez R., Peacock J. 1999, MNRAS, 310, 43  

\bibitem[Wright (2006)]{wright06} Wright E.L. 2006, PASP, 118, 1711



\end{thebibliography}
\end{document}